\documentclass[pra, 
aps,
amsmath,
amssymb,
twocolumn,
nofootinbib,
superscriptaddress,
]{revtex4-2}
\usepackage{graphicx,dcolumn,bm}
\usepackage{amsfonts,amsmath,amssymb,amsthm,amscd}
\usepackage[english]{babel}
\usepackage{physics}
\usepackage{needspace}
\usepackage{xcolor}
\usepackage{bm}
\usepackage{slashed}
\usepackage[export]{adjustbox}
\usepackage{enumerate}
\usepackage{tabularx}
\usepackage{tablefootnote}
\usepackage{ulem}

\definecolor{linkblue}{HTML}{2e3092}
\usepackage[colorlinks=true, allcolors = linkblue]{hyperref}
\renewcommand{\vec}[1]{\bm{#1}}

\AtBeginDocument{\renewcommand{\natexlab}[1]{#1}}

\allowdisplaybreaks

\begin{document}
\title{Strong-field ionization in particle-in-cell simulations}
\author{A.A. Mironov}
\email[Correspondence email address: ]{mironov.hep@gmail.com }
\affiliation{Center for Theoretical Physics (CPHT), CNRS, \'{E}cole Polytechnique, Institut Polytechnique de Paris, 91128 Palaiseau, France}
\author{E.G. Gelfer}
\affiliation{ELI Beamlines Facility, The Extreme Light Infrastructure ERIC, 25241 Dolni Brezany, Czech Republic}
\affiliation{HiLASE Centre, Institute of Physics of the Czech Academy of Sciences, 25241, Dolni Brezany, Czech Republic}
\author{I.I. Tupitsyn}
\affiliation{Department of Physics, St. Petersburg State University, Universitetskaya 7-9, 199034 St. Petersburg, Russia}
\author{A. Beck}
\affiliation{Laboratoire Leprince-Ringuet – \'{E}cole Polytechnique, CNRS-IN2P3, Palaiseau 91128, France}
\author{M. Jirka}
\affiliation{ELI Beamlines Facility, The Extreme Light Infrastructure ERIC, 25241 Dolni Brezany, Czech Republic}
\affiliation{FNSPE, Czech Technical University in Prague, Prague, Czech Republic}
\author{O. Klimo}
\affiliation{ELI Beamlines Facility, The Extreme Light Infrastructure ERIC, 25241 Dolni Brezany, Czech Republic}
\affiliation{FNSPE, Czech Technical University in Prague, Prague, Czech Republic}
\author{S. Meuren}
\affiliation{LULI, CNRS, CEA, Sorbonne Universit\'{e}, École Polytechnique, Institut Polytechnique de Paris, F-91128 Palaiseau, France}
\author{G. Oberreit}
\affiliation{Laboratoire Leprince-Ringuet – \'{E}cole Polytechnique, CNRS-IN2P3, Palaiseau 91128, France}
\author{T. Smorodnikova}
\affiliation{Stanford PULSE Institute, SLAC National Accelerator Laboratory, Menlo Park, CA 94025, USA}
\author{R. Ta\"ieb}
\affiliation{Sorbonne Universit\'{e}, CNRS, Laboratoire de Chimie Physique-Mati\`{e}re et Rayonnement, LCPMR, F-75005 Paris, France}
\author{S. Weber}    
\affiliation{ELI Beamlines Facility, The Extreme Light Infrastructure ERIC, 25241 Dolni Brezany, Czech Republic}
\author{C. Riconda}
\affiliation{LULI, Sorbonne Université, CNRS, CEA, École Polytechnique, Institut Polytechnique de Paris, F-75255 Paris, France}
\author{M. Grech}
\affiliation{LULI, CNRS, CEA, Sorbonne Universit\'{e}, École Polytechnique, Institut Polytechnique de Paris, F-91128 Palaiseau, France}
\author{S.V. Popruzhenko}
    \affiliation{National Research Nuclear University MEPhI, Kashirskoe shosse 31, 115409, Moscow, Russia}
    \affiliation{Prokhorov General Physics Institute RAS, Vavilova 38, 119991, Moscow, Russia}

\begin{abstract}
The inclusion of the process of multiple ionization of atoms in high-intensity electromagnetic fields into particle-in-cell (PIC) codes applied to the simulation of laser-plasma interactions is a challenging task. In this paper, we first revisit ionization rates as given by the Perelomov-Popov-Terent'yev formulas within the paradigm of sequential tunnel ionization. We analyze the limit of validity and possible inconsistencies of this approach. We show that a strongly limiting factor to a precise description of ionization is the competing contribution of different sequential ionization processes. To solve this an algorithm is proposed that allows to find the dominant nonsequential path of tunnel ionization, and significantly improves the precision in simulations. This novel procedure is implemented in the PIC code SMILE, and includes the dependence of the ionization rates on the magnetic quantum number of the level. The sensitivity to variations in the ionization model is studied via full simulations of  the ionization of an argon target by an incident high-intensity laser pulse. Finally,  we analyze generalizations of the Perelomov-Popov-Terent'yev rate developed to describe the barrier suppression ionization in high fields and discuss the necessity and possibility of including these extensions in PIC simulations.
\end{abstract}

\maketitle

\section{Introduction}

In recent decades, particle-in-cell (PIC) simulations have become a pivotal research instrument in the physics of intense electromagnetic fields' interaction with matter \cite{birdsall_book2018,arber_ppcf2015,gonoskov_pre2015}.  
Although the PIC method was initially designed and applied for the numerical description of the dynamics of fully ionized, collisionless plasmas \cite{dawson_rmp1983,pukhov_prl1996,fonseca2013exploiting}, the later prominent progress in the code architecture and the capacity of computer resources has led to a significant expansion of the PIC-based approaches into the physics of many-body classical and quantum systems.
The severe restriction on the evolution time of a system simulated by the PIC method still imposes an upper bound on the level of tens of picoseconds, limiting the scope of consideration by phenomena occurring in the interaction of femto- and attosecond electromagnetic pulses with matter.
At the same time, modern PIC codes account for not only the effects of collective fields self-consistently, but also are capable of taking into consideration different elementary processes, which may happen with individual charged particles \cite{takizuka_jcp1977,birdsall_ieee1991,nanbu_pre1997,perez_pop2012,sentoku_pop1998,vyskocil_ppcf2018,martinez_pop2019,sokolov_pop2009,sironi_aj2009,bussmann_proc2013,duclous_ppcf2010,nerush_prl2011,arber_ppcf2015,gonoskov_pre2015,penetrante_pra1991,chessa_pop1998,bruhwiler_pop2003,kemp_pop2004,fonseca2013exploiting,arber_ppcf2015,nuter_pop2011,nerush_prl2011,gonoskov_pre2015,lobet_jpcs2016,fedeli_warpx2022}.

Here, the characterization ``elementary process'' stands for any event (beyond a particle push by an electromagnetic field) that involves some of the particles, not necessarily all of them truly elementary.
This includes, in particular, binary electron-electron, electron-ion and ion-ion collisions \cite{takizuka_jcp1977,birdsall_ieee1991,nanbu_pre1997,perez_pop2012}, bremsstrahlung \cite{sentoku_pop1998,vyskocil_ppcf2018,martinez_pop2019}, Thomson \cite{sokolov_pop2009,sironi_aj2009,bussmann_proc2013} and Compton radiation \cite{duclous_ppcf2010,nerush_prl2011,arber_ppcf2015,gonoskov_pre2015,lobet_jpcs2016, niel_pre2018, niel_ppcf2018}, Breit-Wheeler process \cite{duclous_ppcf2010,nerush_prl2011,arber_ppcf2015,gonoskov_pre2015,lobet_jpcs2016} and ionization of atoms and ions by impact \cite{perez_pop2012} or under the action of external or self-consistent electromagnetic fields \cite{penetrante_pra1991,chessa_pop1998,bruhwiler_pop2003,kemp_pop2004,fonseca2013exploiting,arber_ppcf2015,nuter_pop2011}.
Cross sections or rates of these processes are known either in the form of exact results or within approximations.
Presently, several PIC codes with all essential elementary processes included are being used for simulations of laser-plasma dynamics and radiation in a broad range of parameters.
For example, the list includes SMILEI \cite{derouillat_cpc2018}, EPOCH \cite{brady_ppcf2010,arber_ppcf2015}, OSIRIS \cite{fonseca2013exploiting}, PICADOR \cite{bastrakov_jcc2012}, PIConGPU \cite{burau_ieee2010}, UMKA \cite{umka-1,liseykina_prl13}, and WARPX \cite{fedeli_warpx2022}.
These codes have already allowed the community to make significant progress in understanding laser-plasma interaction under extreme conditions when it is necessary to include radiation reaction, production of electron-positron pairs, generation of QED cascades, excitation of strong magnetic fields and more, see papers \cite{gonoskov_rmp2022,popruzhenko_ufn2023,fedotov_physrep2023} for the state-of-the-art review.

Difficulties in including elementary processes in PIC simulations stem essentially from the fact that the parameters of plasma and laser fields span very wide intervals.
In particular and most importantly, laser intensity can vary from $10^{15}$ to $10^{25}$W/cm$^2$ and even higher. The lower bound corresponds to the characteristic atomic field, where fast ionization and production of multiply charged ions occur. The upper bound yet remains out of experimental reach and marks the threshold for nonlinear quantum electrodynamic (QED) effects \cite{fedotov_physrep2023, gonoskov_rmp2022, popruzhenko_ufn2023} such as the generation of QED cascades \cite{Kirk2008, fedotov2010limitations}.
In many cases, there is no way to provide an analytic quantitatively correct description of a cross-section or rate, which would be universally valid in the whole aforementioned interval of intensities. Therefore, practical implementations in codes will impose additional restrictions on intensity intervals or other parameters.

In this paper, we analyze in depth the process of strong-field ionization from the viewpoint of its description in PIC codes.\footnote{We use SMILEI \cite{derouillat_cpc2018} as a reference code and suggest its modification in this work.}
A quantitatively correct description of the multiple ionization process of heavy-atomic targets in the simulation of plasma dynamics is important for several reasons.
Firstly, it is required to correctly describe the charge distribution of laser-created plasmas at different stages of the interaction.
The free electron concentration is one of the key parameters determining the plasma response to intense laser radiation. It impacts the plasma formation itself \cite{azamoum_lightsciapp2024}, the build-up of sheath fields \cite{hall_aipadv2023}, the overall plasma expansion dynamics \cite{bernert_scirep2022} or ion acceleration \cite{nuter_pop2011,wang_prx2021}. 
Secondly, multiple ionization of heavy atoms inside a laser pulse of ultra-high intensity can be used as a specific method of electron injection for their subsequent acceleration \cite{mcguffey_prl2010,pak_prl2010,pollock_prl2011,thaury_scirep2015}, or to initiate avalanche-type QED cascades \cite{artemenko_pra2017}.
Thirdly, measuring the maximal charge state of heavy ions can be used to determine the peak intensity in the focus of laser pulses. This method was tested at relatively moderate intensities in Refs.~\cite{walker_pra01,ueda_pra03,link_instr06}. The lack of \textit{in situ} intensity measurement techniques suitable for the future multi-petawatt laser facilities motivated recent theoretical treatment of the heavy-atom ionization in extreme fields  \cite{ciappina_pra2019,ciappina2020focal,ciappina_lpl2020}.
In light of this, the quantitatively correct treatment of field-induced ionization in PIC simulations is in high demand.
Note that here we do not consider collisional ionization, which is definitely significant but parametrically separated from the domain, where strong-field ionization is pivotal. 
With the concentration of atoms growing, the contribution of collisions into the rate of electron transitions in ions become dominant at some threshold value, which depends on the laser intensity, pulse duration, and on the target chemical composition. 
We assume a density of atoms in the gas below this threshold.
For modeling the high-density regime of interaction with the effects of collisions taken into account, see \cite{helios_06,pikuz_06} and references therein.

In available PIC codes, the implementation of the multiple strong-field ionization process is based on a Monte Carlo algorithm. It uses the rates of single-electron ionization described by the analytic formulas of Perelomov, Popov and Terent'ev (PPT) \cite{perelomov_jetp1966,perelomov_jetp1967,perelomov_jetp1967b}, see also reviews \cite{popov_ufn2004,poprz_jpb14}.
These rates also are known in the literature as ADK were rederived by Ammosov, Delone and Krainov in 1986 \cite{ammosov_jetp1986}. Overall, these approximate formulas, although applicable in a wide semiclassical domain, can leap out of their range of validity when the simulation parameters change. We analyze some situations where such invalidation may happen, in particular, in the barrier suppression regime \cite{ciappina_lpl2020}, and compare possible modifications of the PPT rates, suitable for implementation in PIC codes \cite{tong_jphysb2005, kostyukov_pra2018, golovanov2020formula, ouatu_pre2022}. 

It is common to assume in PIC simulations \cite{nuter_pop2011}, that the ionization process is \textit{sequential}, namely, electrons are extracted from the ion one by one in the order of their energy levels starting from the outermost ones. This is a strong approximation, which, however, may seem unavoidable for practical implementations. 
We argue that electrons occupying close levels can be ionized \textit{nonsequentially} at a high probability. 
Accounting for this in PIC simulations can significantly affect the result. 
We propose a way to identify the most probable ionization pathway and suggest an algorithm for including it in the PIC method. We implement it in SMILEI and analyze the accuracy of the sequential approximation. The discussion is illustrated with detailed simulations for the ionization of argon by a short intense laser pulse.

The paper is organized as follows. In Section~\ref{sec:ii}, we discuss the fundamental constraints on the implementations of strong-field ionization in PIC codes. Section~\ref{sec:iii} is dedicated to the review of specific analytical strong-field ionization models and their accuracy analysis. Then, in Section~\ref{sec:iv}, we discuss the significance of the nonsequential strong-field ionization on the example of argon. We suggest a new scheme for consistent treatment of the ionization rate dependence on the magnetic quantum number and account for the nonsequential ionization pathway in the PIC method. In Section~\ref{sec:v}, we study in detail the interaction of a short laser pulse with an argon target. We test our new scheme against the solution to the system of rate equations, then evaluate the error imposed by the sequential ionization approximation, and, finally, study the impact of the barrier suppression modification of the tunneling formula on the PIC simulations outcome. Section~\ref{sec:vi} provides the conclusion and outlook.

\section{Fundamental constraints on a universal description of multiple strong field ionization in PIC codes}
\label{sec:ii}

We start from a brief discussion of the most significant constraints commonly imposed on the description of nonlinear strong-field ionization in PIC simulations. 
The two widely used limitations are:
\begin{enumerate}
    \item[I.] The ionization process is assumed of a single-electron nature, without any significant contribution of correlation effects. This means that the electrons are being removed from an atom independently one from another so that the only relevant channel is the production of ions $A^{Z+1}$ from ions $A^Z$, where $Z$ stands for the ion charge state, $Z=0$ for neutral atoms.
    \item[II.] For a given charge state, the outermost electron having the lowest ionization potential $I_p$ detaches with the dominant probability, compared to all other electrons, so that it is sufficient to account for only this channel.
\end{enumerate}

Anticipating our discussion in the following sections, let us emphasise that in this work we extensively revise condition II (see Section~\ref{sec:ivd}). We argue that the sequential order of ionization can be replaced by a nonsequential one. By choosing a correct dominant ionization pathway, the precision of PIC simulations can be improved without affecting the calculation efficiency.

Both assumptions I and II are well justified \textit{only} for (i) multiquantum ionization in a field of (ii) sub-relativistic and higher intensity.
Condition (i) means that the photon energy of the laser field is small compared to the ionization potential,
\begin{equation}
  \hbar\omega\ll I_p~.
  \label{K0}
\end{equation}
This is satisfied for ionization of all atoms in infrared fields of wavelength $\lambda\simeq 1\mu{\rm m}$; for highly charged ions with $I_p$ of several dozens eV and higher, it is amply fulfilled.
Condition (ii) stems from the requirement that the  recollision \cite{corkum_prl93, ivanov_rmp06, krausz_nobel24}, which may happen upon the photoelectron return to its parent ion, is suppressed. 
This suppression results from the lateral relativistic drift.
The drift exceeds the lateral width of the photoelectron wave packet and therefore eliminates the effect of recollision for \cite{keitel_prl17}
\begin{equation}
    a_0> \bigg(\frac{p_{\rm ch}}{mc}\bigg)^{1/2}\bigg(\frac{E_0}{E_{\rm ch}}\bigg)^{1/4}~.
    \label{drift}
\end{equation}
Here,
\begin{equation}
    a_0=\frac{eE_0}{mc\omega}
    \label{a0}
\end{equation}
is the celebrated dimensionless invariant field, $E_0$ is the electric field amplitude of the laser wave of frequency $\omega$, $e$ and $m$ are the elementary charge and electron mass correspondingly, $c$ is the speed of light.
Finally, $p_{\rm ch}$ and $E_{\rm ch}$ are the characteristic electron momentum and electric field of the bound state.

When Eq.~(\ref{K0}) is violated, particularly in the domain $\hbar\omega\simeq I_p$, the few-photon or single-photon ionization cross sections become generally complicated functions of the laser frequency and may contain resonances.
The ionization probability is no longer a monotonically decreasing function of $I_p$ so assumption II does not hold.
In the high-frequency domain, $\hbar\omega\gg I_p$, several channels can compete including ionization of inner shells, ionization with excitation of the residual core and correlated double or multiple ionization by a single photon.
Thus, assumption I also appears invalid.

When Eq.~(\ref{drift}) does not apply, an electron removed from an atom through the single-electron ionization process, can return to this atom, provided the field is polarized linearly (or close to linearly).
This may cause recollision, a pivotal effect of nonlinear atomic optics responsible for the nonsequential double and multiple ionization \cite{kuchiev_jetpl87}, the generation of high order harmonics \cite{corkum_prl93} and the high energy plateau in the photoelectron spectra \cite{paulus_prl94}.
For a comprehensive review of the recollision physics, we refer the reader to Refs.~\cite{ivanov_rmp06, becker_rmp12, paulus_jpb18,agostini_nobel24,huillier_nobel24,krausz_nobel24}.
For our subject, it is significant that the recollision can enhance the ionization probability of the second and the third electron by several orders in magnitude, the effect fully resulting from electron-electron correlations.
In this situation, the rate of doubly and triply charged ion production cannot be described within a single-electron sequential model.
For $\sim 1\,\mu{\rm m}$ wavelength light, condition (\ref{drift}) is satisfied for laser intensity ${\cal I}>10^{16}{\rm W/cm}^2$.

Summarizing, ionization can be introduced in PIC codes within a simple concept of a sequential single-electron process only for low-frequency fields [see Eq.~(\ref{K0})] at intensities higher than $10^{16}-10^{17}{\rm W/cm}^2$ [see Eq.~(\ref{drift})]. 
These requirements are satisfied in most cases, where the PIC method is used to study the dynamics of a laser-plasma interaction.
However, one has to note that, when an atomic target is impinged by a strong laser pulse with a smooth temporal envelope, the ionization of several outermost electrons happens at the early stage of interaction before the pulse peak arrives at the target, namely, at relatively low fields ${\cal I}<10^{16}{\rm W/cm}^2$.
Therefore, it is hard to expect that this stage of the plasma formation will be correctly described in a PIC simulation.
Fortunately enough, as we show below, the formation of \textit{highly charged} ions in intense fields is rather insensitive to the initial condition so that an incorrectly described initial stage of ionization will hardly influence the subsequent dynamics. Finally, the inapplicability of the approach for describing ionization by high-frequency light means that the effect of secondary XUV and X-ray radiation emitted by the plasma on the production of ions remains entirely discarded.

\section{Analytic models for sequential strong field ionization}
\label{sec:iii}
In this section and below, we use atomic units $e=m=\hbar=1$ customary in nonlinear atomic optics and physics of strong-field ionization.
In these units, the speed of light $c=1/\alpha_e\approx 137$ with $\alpha_e$ being the fine structure constant.
In atomic units, the electric and magnetic field strengths are scaled by the atomic electric field $E_{\rm at}=m^2e^5/\hbar^4=5.14\times 10^9{\rm V/cm}$ and ionization potentials --- by the atomic energy $me^4/\hbar^2=27.2{\rm eV}$.
The ionization potential of atomic hydrogen is then $I_p({\rm H})=1/2$.
Laser intensity is expressed in the form
\begin{equation}
    \begin{gathered}
        \mathcal{I}=E^2(1+\rho^2)\mathcal{I}_{\rm at},\\ \mathcal{I}_{\rm at}=\frac{c}{8\pi}\frac{m^4e^{10}}{\hbar^8}=3.51\times 10^{16}{\rm W/cm}^2~,
    \end{gathered}
    \label{I}
\end{equation}
where $E$ is the electric field amplitude in atomic units and $\rho$ is the ellipticity, $\rho=0$ for linear and $\rho=\pm 1$ for circular polarization.

\subsection{Tunneling ionization rates}
\label{sec:iiia}
Under condition (\ref{K0}), ionization of atoms and positive ions proceeds in a highly nonlinear regime, which was theoretically described within an analytic semi-classical approach pioneered by Keldysh \cite{keldysh_jetp1965} in the mid-60s \cite{nikishov_jetp66,perelomov_jetp1966}.
This approach also widely referred to as the Strong Field Approximation (SFA) was later developed in great detail \cite{faisal_jpb73,reiss_pra80}.
For a comprehensive review of the Keldysh theory and SFA, we refer the reader to Refs.~\cite{popov_ufn2004,poprz_jpb14}.

Within the nonlinear regime, the character of ionization dynamics is determined by the value of the Keldysh parameter:
\begin{equation}\label{gamma}
    \gamma=\frac{\sqrt{2I_p}\omega}{E_0}~,
\end{equation}
with $E_0$ being the electric field amplitude of the laser wave.
For $\gamma\ll 1$, ionization proceeds in the tunneling regime, and the laser field can be treated as quasi-static during the process.
In the opposite multiphoton limit $\gamma\gg 1$, ionization can be treated as penetration of the electron through a rapidly oscillating barrier. 
In most cases,  the ionization of higher charge states (e.g. $Z\geq 3$) in optical and infrared fields always proceeds in the regime of tunneling $\gamma\ll 1$ so that the quasi-static field approximation can be safely used for calculating ionization rates.

In this limit, for an ion with charge $Z$, the rate of ionization of a level with orbital and magnetic quantum numbers $l$ and $m$ is given by 
\begin{equation}\label{PPT_definition}
    \begin{gathered}
        \begin{split}
        w^{\rm {PPT}}(t)=&4 C^2_{n^*l} B_{l|m|} I_p\\
        &\times \left( \frac{2}{F(t)} \right)^{2n^*-|m|-1} \exp\left\lbrace -  \frac{2}{3F(t)}\right\rbrace,
        \end{split}\\[1ex]
	B_{l|m|} = \frac{(2l+1)\Gamma(l+|m|+1)}{2^{|m|}\Gamma(|m|+1)\Gamma(l-|m|+1)}.
    \end{gathered}
\end{equation}
Here $n^*=Z^*/\sqrt{2I_p}$ is the effective principal quantum number, $Z^*=Z+1$ is the atomic residual charge,
$Z^*\leq Z_{\rm m}$, where $Z_{\rm m}$ is the atomic number, and $\Gamma(x)$ is the Euler Gamma function.
The dimensionless value 
\begin{equation}
    F(t)=\frac{E(t)}{E_{\rm ch}}~,~~~E_{\rm ch}=(2I_p)^{3/2}\equiv\bigg(\frac{I_p}{I_p(H)}\bigg)^{3/2}
    \label{F}
\end{equation}
called \textit{reduced electric field} is the ratio of the absolute value of the laser electric field to the characteristic field of the atomic level $E_{\rm ch}$.
Finally, $C_{n^*l}$ is the asymptotic coefficient of the single-electron wave function of the level at large distances from the ionic core \cite{popov_ufn2004}:
\begin{equation}
    \psi_{n^*lm}({\bf r})\approx 2C_{n^*l}\sqrt{\kappa}(\kappa r)^{n^*-1}e^{-\kappa r}Y_{lm}\bigg(\frac{{\bf r}}{r}\bigg)~,
    \label{psi}
\end{equation}
where $\kappa=\sqrt{2I_p}$ is the characteristic momentum of the bound state and $Y_{lm}(\mathbf{r}/r)$ are the spherical harmonics. The rate (\ref{PPT_definition}) was derived by Perelomov, Popov and Terent'ev in 1966-67 \cite{perelomov_jetp1966,perelomov_jetp1967,perelomov_jetp1967b,popov_ufn2004} and is now universally referred to as the PPT rate.
The derivations and comparisons with the experimental data and numerical results can be found in reviews \cite{popov_ufn2004,poprz_jpb14} and references therein.

Equation (\ref{PPT_definition}) is an approximation valid under the following conditions:\\
\textbf{(a)} Low-frequency field $\gamma\ll 1$ (\ref{gamma}) and the multiquantum regime, see Eq.~(\ref{K0}).
Both requirements are amply satisfied for ionization in strong optical or infrared fields for intensities above $10^{16}{\rm W/cm}^2$.\\
\textbf{(b)} Nonrelativistic approximation for the description of the tunneling effect. 
Although dynamics of free electrons becomes relativistic at $a_0\simeq 1$, and this leads to substantial relativistic effects in photoelectron spectra already for ${\cal I}\simeq 10^{17}{\rm W/cm}^2$, the total ionization rate remains determined by nonrelativistic physics under the condition \cite{popov_ufn2004, ciappina_pra2019}
    \begin{equation}\label{nonrel}
        I_p\ll mc^2\simeq 137^2.
    \end{equation}
Here, we retain the electron mass $m$ for clarity.
This condition means that the bound state is nonrelativistic.
According to estimates made in Ref.~\cite{ciappina_pra2019}, the condition remains valid for the calculation of ionization rates up to intensities ${\cal I}\simeq 10^{25}{\rm W/cm}^2$. 
One should note here that although the condition (\ref{nonrel}) does not contain laser parameters, we can ultimately derive a restriction on the laser intensity from there.
This can be explained by taking into account that for a given value of $I_p$ ionization takes place at certain amplitudes of the reduced field (\ref{F}), typically $F=0.03..0.05$, when the rate (\ref{PPT_definition}) becomes sufficiently large.
More precisely, this estimate for the reduced field $F$ was made in \cite{ciappina_pra2019} assuming that the probability of ionization calculated with the rate (\ref{PPT_definition}) with $m=0$ becomes close to unity in a single laser period $2\pi/\omega$. 
This allows to define the threshold electric field $E_{\rm th}\simeq (0.03..0.05)(2I_p)^{3/2}$ required to ionize the level with the given $I_p$.
Taking $I_p=0.2mc^2$ and 0.05 for the coefficient in $E_{\rm th}$, one obtains the above estimate for the laser intensity.

\noindent
\textbf{(c)} The semi-classical approximation used to calculate the rate (\ref{PPT_definition}) applies when the electron energy $-I_p$ is well below the maximum of the potential barrier created by joint action of the Coulomb and the laser electric fields. 
In this case, the electron tunnels through a wide barrier, such that its width $b\approx I_p/E(t)$ is much higher than the De Broglie wavelength.
This implies
\cite{popov_ufn2004,tong_jphysb2005,ciappina_lpl2020,kostyukov_pra2018,ouatu_pre2022}:
    \begin{equation}\label{BSIcond}
        E\ll E_{\rm BS}=\frac{I_p^2}{4Z^*}.
    \end{equation} 
Depending on the laser pulse duration and the values of $n^*$, $l$, and $m$ numbers of the bound state, the inequality (\ref{BSIcond}) can be violated before the corresponding level is fully ionized (see the discussion and examples in Ref.~\cite{ciappina_lpl2020}).
This circumstance dictates the necessity to significantly correct the semiclassical PPT rate in the high-field domain.
Some models developed to extend the semiclassical description to the barrier-suppression ionization (BSI) regime are discussed in the next subsection.\\
\textbf{(d)} Finally, the coefficient $C_{n^*l}$ in Eqs.~(\ref{PPT_definition}), (\ref{psi}) can either be found numerically within the Hartree-Fock or other self-consistent field approach \cite{radzig_book2012, vitanov1991asymptotic} or estimated using the asymptotic formula suggested by Hartree \cite{hartree}:
\begin{equation}
    C^2_{n^*l} =  \frac{2^{2n^*-2}}{n^*\Gamma(n^*+l+1)\Gamma(n^*-l)}.
\label{hartree}
\end{equation}
This approximation works generally well; see comparisons of this coefficient calculated by different methods for argon in Appendix~\ref{app:cnl}.
However, Eq.~(\ref{hartree}) may fail for some neutral states ($Z=0$) as well as for $p$, $d$ and higher $l$ states in ions with small $Z$. 
In these cases, one may set $C_{n^*l}=1$ without losing the overall accuracy (see also Appendix~\ref{app:cnl}).

In the tunnel ionization theory known in the literature as ADK \cite{ammosov_jetp1986}, the orbital quantum number $l$ in Eq.~(\ref{hartree}) is replaced by the effective orbital number $l^*=n^*-1$ (in the ground state).
In the context of the ionization problem, this replacement (which solely distinguishes the ADK rate from the PPT one) does not seem to be justified. 
Still this has some minor effect on the accuracy of the PPT formula.

\subsection{BSI regime}
\label{sec:iiib}
With the electric field of the laser wave approaching the value $E_{\rm BS}$ (\ref{BSIcond}), the PPT formula (\ref{PPT_definition}) becomes progressively inaccurate, strongly overestimating the rate \cite{popov_ufn2004}.
Condition (\ref{BSIcond}) can be rewritten in terms of the reduced field (\ref{F})
\begin{equation}
    F\ll \frac{\sqrt{2I_p}}{16Z^*}=\frac{1}{16n^*}~,
    \label{FBSI}
\end{equation}
which shows that the applicability of the tunnel approximation breaks earlier for states with higher effective quantum numbers (see also the next subsection).

Different models have been suggested to estimate the rate of ionization in the domains $E\simeq E_{\rm BS}$, $E>E_{\rm BS}$ and $E\gg E_{\rm BS}$.
Let us emphasize two significant points.
Firstly, in contrast to the semiclassical quasi-static domain specified by
\begin{equation}
    F\ll 1~,~~~\gamma\ll 1~,
    \label{Fgam}
\end{equation}
in the BSI regime, no analytic methods exist, that allow for calculating the rate in a closed form (see e.g. \cite{popov_ufn2004} for further information).
Thus, all analytic formulas for the rate known for this regime, although based on some physics models, should be considered as extrapolation.
Secondly, with the reduced field $F$ increase, the ionization rate grows fast, so that only the case $F<0.2..0.3$ is of practical interest.
If the laser intensity is so high that larger values of $F$ can be achieved, the respective level will be stripped of electrons completely before the field maximum arrives at the target.
The respective estimates can be found in Ref.~\cite{ciappina_pra2019}.
Thus, for calculation of ion yields in intense laser fields, one predominantly needs to use, apart from the main expression (\ref{PPT_definition}), its suitable extension into the domain of higher field, up to $E\simeq E_{\rm BS}$, while the regime $E\gg E_{\rm BS}$ is usually not of practical interest. 

Several methods are known for extrapolating the semiclassical ionization rates in the strong-field domain \cite{popov_ufn2004, tong_jphysb2005,bauer_pra1999,zhang_pra2014}.
The one suggested by Tong and Lin (TL) in \cite{tong_jphysb2005} employs an empirical factor which suppresses the PPT rate with $E$ approaching $E_{\rm BS}$:
\begin{equation}\label{rate_TL}
        w^{\rm TL} = w^{\rm PPT} \exp\left(-\frac{\alpha}{8}\frac{E}{E_{\rm BS}} n^*\right),
	\end{equation}
where $\alpha$ is a constant adjusted to fit the numerically calculated rate.
For examples considered in \cite{tong_jphysb2005} the best fit was achieved for hydrogen in the interval $\alpha=6..9$, meaning that $\alpha n^*/8\simeq 1..2$. 
The extrapolation given by Eq.~(\ref{rate_TL}) works well for $E/E_{\rm BS}=0.1..0.5$ \cite{tong_jphysb2005,ciappina_lpl2020}, but strongly underestimates the rate for $E>E_{\rm BS}$ \cite{kostyukov_pra2018}. 

Kostyukov, Artemenko and Golovanov (KAG) suggested an alternative piecewise approximation of the ionization rate \cite{golovanov2020formula} that joins the PPT rate \eqref{PPT_definition} with the approximate BSI formulas derived by Bauer and Mulser for $E\sim E_{\rm BS}$ \cite{bauer_pra1999} and by KAG for $E\gg E_{\rm BS}$ \cite{artemenko_pra2017,kostyukov_pra2018}. The suggested formula
   \begin{equation}\label{rates_KAG}
         w^{\rm KAG}=
         \left\lbrace
        \begin{array}{ll}
            w^{\rm PPT}(E)~, & E<E_1~,\\
            \alpha_1 E^2/I_p^2~, &  E_1\leq E<E_2~,\\
            \alpha_2 E/I_p~, &  E\geq E_2~,
        \end{array} \right.
   \end{equation}
extrapolates the rate to considerably stronger fields.
Here the coefficients $\alpha_{1,2}\sim 1$. Following Refs.~\cite{golovanov2020formula,bauer_pra1999}, we set $\alpha_1=2.4$ and $\alpha_2=0.8$. The fields $E_{1,2}$ are defined by requiring the continuity of the piecewise formula \eqref{rates_KAG}. 
The expression for $E>E_2$ is obtained by assuming that the laser field is much stronger than the atomic field and neglecting the electron kinetic energy with respect to its energy in the laser field. The intermediate Bauer-Mulser formula serves as a link between the PPT rate and the high-field KAG solution.

The KAG approach was implemented in the SMILEI PIC code by Ouatu et al \cite{ouatu_pre2022}. We also introduce this model in our version of the code \cite{github}, as well as the TL correction, which allows us for their direct comparison in Section~\ref{sec:v}.

\subsection{Hierarchy of quantitative inaccuracies}
\label{sec:iiic}
In this subsection, we briefly describe the main flaws and potential sources of inaccuracies of the common scheme for including strong-field ionization in PIC codes.
In Section~\ref{sec:v} and Appendix~\ref{app:cnl} these flaws are examined numerically to demonstrate their impact.

\subsubsection{Coefficients $C_{n^* l}$}
\label{sec:iiic1}
The coefficients $C_{n^* l}$ in Eqs.~(\ref{PPT_definition}) and (\ref{psi}) can be found numerically within the Hartree-Fock or other appropriate model for many-electron atoms, while the asymptotic formula (\ref{hartree}) and its modification used in Ref.~\cite{ammosov_jetp1986} may leap out of the applicability range for some parameter values.
The difference between the coefficients found numerically and the analytic asymptotic is demonstrated in Appendix~\ref{app:cnl}. 
The accuracy of Eq.~(\ref{hartree}) improves with the increase of the effective quantum number $n^*$. 
For several outer levels, the difference in the factor $C_{n^*l}^2$ can reach almost an order of magnitude.
The safest way to reduce this source of errors is to use numerically calculated coefficients, but this requires quite considerable work to find them for all charge states of all elements.
However, as we argued in Section~\ref{sec:ii}, ionization of several outermost electrons can proceed via the correlated re-colission mechanism, so it is hard to expect a correct result within the chosen model of ionization anyway. As ionization of these levels happens at relatively low intensities, this stage of the process is of marginal interest for most of the cases of laser-plasma interactions at high intensities simulated with the PIC method.\footnote{Below we demonstrate that the ionization of inner electrons is almost independent of the details of ionization of the outermost ones. Therefore, in practice it is convenient to initialize the simulation with already ionized plasma $Z>1$.}

In effect, using the Hartree asymptotic (\ref{hartree}) still provides good quantitative accuracy for most states,
and the approximation suggested in \cite{ammosov_jetp1986} has no advantages for such calculations. 
We provide a version of the open-source code SMILEI \cite{github} that implements the coefficients $C_{n^*l}^2$ in the approximation (\ref{hartree}), or, alternatively, allows using the tabulated values for $C_{n^*l}^2$, for example, calculated by evaluating numerically asymptotic single-electron wavefunctions \cite{vitanov1991asymptotic}.
    
\subsubsection{Dependence on the magnetic quantum number $m$}
\label{sec:iiic2}
The PPT rate in Eq.~(\ref{PPT_definition}) depends on the magnetic number $m$ with the quantization axis in the polarization direction of the electric field.
The main $|m|$-dependent factor $(2/F)^{-|m|}$ stems from the fact that for $m\ne 0$ the bound state wave function is zero in the polarization direction.
Since fast tunnel ionization happens at $F=0.03..0.05$ \cite{ciappina_pra2019}, the additional suppression factor is $\sim 10^{-2}$ for $p$-states with $m=\pm 1$ and $\sim 10^{-4}$ for $d$-states with $m=\pm2$.
In other words, levels with $m=0$ ionize much faster than the others.
The $m$-connected ambiguity of the calculation comes from the fact that different models for the population of the magnetic sublevels can be applied.
In the approach for PIC simulations, initially suggested by Nuter et al \cite{nuter_pop2011}, it is assumed that all present electrons of $p$-, $d$- and higher subshells are distributed in $m$ with equal probabilities.
In this case the rate of ionization from an $l$-shell populated by $N\le 4l+2$ electrons in a $Z$-charged ion is\footnote{We note that in the original work \cite{nuter_pop2011} the factor $N/(2l+1)$ is not present.  While it is typically $\sim 1$ (for $p$-electrons, it ranges from 1/3 to 2), it should be included for higher precision.}
\begin{equation}\label{rate_m_zero}
    \begin{split}
    w(Z,I_p,E(t),l)&=\frac{N}{2l+1}\sum_m w(Z,I_p,E(t),l,m)\\
    &\approx \frac{N}{2l+1}w(Z,I_p,E(t),l,m=0)~,     
    \end{split}
\end{equation}
where we used shorthand notation $w$ for the ionization rate as a function of the listed parameters.
Here, we took into account that for $m=\pm 1$ and higher the probability of ionization is only a few per cent of that for $m=0$.
This model assumes the presence of an interaction which redistributes all the remaining electrons homogeneously in $m$ after each ionization step.
We do not see well-justified arguments on what this interaction could be (at least for low densities when collisions are negligible).
Instead, some arguments can be provided supporting conservation of $m$.
Firstly, with the ion charge $Z$ increasing the effective value of the Keldysh parameter (\ref{gamma}) decreases, and for the levels of interest $\gamma\approx 10^{-1}$ and less.
This means that ionization proceeds as in a static field, and its slow time variation can be considered adiabatic, which makes the field-free quantum numbers well conserved.
Secondly, the difference in ionization rates for sub-levels with different $m$ was demonstrated in experiments on strong field ionization of dilute gases (so that collisions were negligible), which was also used to control photoelectron circular currents, see, e.g. \cite{smirnova_np18}. Dynamics of $m$-dependent tunneling was found generally in agreement with theoretical predictions \cite{smirnova_pra11,barth_jpb14}.

As an alternative convenient for implementation in PIC codes, we suggest assuming that the $m$ number is conserved. 
This can be justified to some extent by the fact that the magnetic field of the laser wave splits the sublevels slightly so that their energies are not exactly equal and, in the absence of other interactions, no transitions between sublevels with different $m$ happen. 

Taking into account the $m$-suppression in the probabilities, one may simply assume that two electrons with $m=0$ are being removed first, then sublevels with $m=\pm 1$ are being ionized, etc. We implement this new procedure in SMILEI \cite{github}. In Section~\ref{sec:v}, we illustrate our approach to treating the dependence on $m$ and compare it to the model suggested in Ref.~\cite{nuter_pop2011}. 

Let us emphasize that the account of nonzero $m$ numbers immediately introduces another ambiguity for levels with close ionization potentials, as inner-shell electrons with $m=0$ can have a higher probability of being extracted than the outer ones with higher $|m|$.

\subsubsection{BSI regime}
\label{sec:iiic3}
With the growth of the laser field strength relative to the barrier suppression field $\sim E_{\rm BS}$ [see Eq.~(\ref{BSIcond})], the semiclassical formula (\ref{PPT_definition}) becomes less accurate as it has been discussed in the previous subsection.
This flaw is the most difficult to fix with controllable accuracy (see discussion in the previous subsection). 
However, the situation remains bearable mostly because it is difficult to reach high values of $E/E_{\rm BS}$ (or $F$) before the level is entirely ionized.
Following the estimate from Ref.~\cite{ciappina_pra2019}, the reduced field required to ionize a level with $m=0$ within a single optical period amounts to $F_0=0.03..0.05$. 
    Then, at this threshold
    \begin{equation}
    \frac{E}{E_{\rm BS}}=4F_0n^*=(0.1–0.2)n^*~,
    \end{equation}
which shows that the tunneling formula is less accurate for larger $n^*$. 
In particular, deviations from the semi-classical behavior and consequent numerical errors are higher for $p$ and $d$ states than for $s$ states \cite{ciappina_lpl2020}.

As we already mentioned, there is no single solution to cure this source of errors, as we are not aware of a rigorously derived analytical expression for the BSI ionization rates, which are universal below and above $E_{\rm BS}$. Expressions \eqref{rate_TL} and \eqref{rates_KAG} are two examples of an approximate solution with a limited applicability range. We use them in this work for a qualitative study. Both depend on external parameters, which can be identified empirically or from a numerical solution to the Schrodinger equation, as, for example, was suggested for the TL model \cite{tong_jphysb2005}.

Let us note that the generally considerable deviation of the modified rate (\ref{rates_KAG}) from the PPT can make no effect in the case of a laser pulse with a smooth envelope impinging on a target. 
Most ionization events happen then at lower fields before the KAG asymptotic $E\gg E_{\rm BS}$ is reached. We illustrate this in Section~\ref{sec:v}.

\subsubsection{Successive ionization of levels with close ionization potentials}
\label{sec:iiic4}
Finally, the adopted scheme discards the possibility that levels with higher ionization potentials can be ionized before those with lower ionization potentials (see the second fundamental constraint in Section~\ref{sec:ii}).
In the semiclassical domain, this assumption is generally well justified for the rate (\ref{PPT_definition}) as it drops exponentially with increasing $I_p$.
Exceptions may happen when two ionization potentials are close in value.

As an example, consider the ionization of argon, the element we use to benchmark our calculations in this paper.
In Ref.~\cite{saloman-argon}, the energy levels and spectra lines of argon are collected for all ionic states up to ${\rm Ar}^{17+}$.
Here we can find that one of the smallest relative differences in ionization potentials is achieved in ${\rm Ar}^{13+}$. 
While $I_p=754$ eV for the outermost $p$-electron in the configuration $1s^22s^22p$, to remove a $2s$ electron from the same configuration one needs $\Delta I_p\approx 50$eV more energy. 
Taking that ionization proceeds efficiently for $F\approx F_0=0.03..0.05$, we estimate that the probability to remove a $2s$ electron differs from that for the outermost $2p$ with $m=0$ by factor $\exp\{-\Delta I_p/(I_pF_0)\}=0.11..0.26$.
On the other hand, for a $p$-electron with $|m|=1$, additional factor $F^{|m|}\approx F_0$ emerges.
As a result, the rate for the inner $s$-electron may appear higher, breaking down the sequential process.\footnote{One should note that for electrons extracted in a nonsequential order, the ionization potentials change. They can be calculated from spectroscopy data. For our argon calculations, we used the data of Ref.~\cite{saloman-argon}.} 
Thus, the discard of multiple or alternative ionization pathways can introduce a greater error than three other factors described above. 

As we will show below the accuracy of the sequential ionization approximation can be on the level of $20-30\%$ or even lower for ionization of levels with close $I_p$-s, such as $2p-2s$, $3p-3s$, $4p-3d-4s$, etc.  
In contrast, for states separated from the higher orbitals by a broad gap (e.g., for $2p^6$ or $1s^2$ electrons), the threshold value of the field required for complete ionization of such levels is not sensitive to the pathway choice.

Generally, a rigorous solution to the issue requires accounting for multiple ionization pathways, which is intricate to implement in the PIC scheme. However, if a dominant pathway can be identified, a refined implementation is possible by replacing the sequential order of ionization with this pathway.

\section{Numerical implementations}
\label{sec:iv}
Consider a rarefied gas of atoms and ions interacting with an incident high-intensity laser pulse of finite duration. 
We assume that the high-intensity tunneling regime $\gamma\ll 1$ is quickly achieved so that the rate Eq.~\eqref{PPT_definition} is applicable.
In order to describe the ionization dynamics quantitatively, one has to apply a numerical procedure. 
We discuss two approaches: incorporation of strong-field ionization into PIC codes \cite{penetrante_pra1991,chessa_pop1998,bruhwiler_pop2003,kemp_pop2004,fonseca2013exploiting,arber_ppcf2015,nuter_pop2011} and implementation of rate equations \cite{ciappina_pra2019,ciappina_lpl2020}. 

\subsection{PIC simulations}
\label{sec:iva}
Ionization can be included in the PIC simulation loop \cite{penetrante_pra1991,chessa_pop1998,bruhwiler_pop2003,kemp_pop2004,fonseca2013exploiting,arber_ppcf2015,nuter_pop2011, chen2013numerical}. At each timestep, the Monte Carlo procedure is applied for each macroparticle with charge $Z$ ($0\leq Z<Z_{\rm m}$). If ionization occurs, a macroparticle is assigned a new charge $Z+1$, a macroelectron is added to the simulation, and the electromagnetic field loses energy \cite{rae_pra1992,nuter_pop2011} spent on ionization. The timestep typically amounts to a fraction of the field period for better performance of the PIC scheme. However, it can be still large when compared to the mean lifetime of one or more electronic bound states.
The algorithm suggested by Nuter et al. \cite{nuter_pop2011, derouillat_cpc2018} resolves this issue by accounting for the possibility of several consecutive ionization events taking place within one timestep. 

A great advantage of the PIC approach is the possibility to study the effect of ionization on laser-plasma interaction, such as the injection of electrons for laser-plasma acceleration schemes \cite{nuter_pop2011}, or interaction with neutral heavy atomic gas targets \cite{ouatu_pre2022}.

The algorithm of Nuter et al \cite{nuter_pop2011} is based on the sequential outermost electron extraction hypothesis and uses the tunneling ionization rates \eqref{PPT_definition}. The original implementation also relies on an additional simplification setting the magnetic quantum number $m$ to zero for all electrons in the view of $w^{\rm PPT}|_{m=0}\gg w^{\rm PPT}|_{m\neq0}$. However, as we discussed above, this does not seem well justified, especially for rare plasmas. We discuss the ways to bypass this approximation in Section~\ref{sec:ivc}. 
Another auxiliary assumption suggests using the expression for the Hartree coefficient $C_{n^*l}$ [see Eq.~\eqref{hartree}] with $l$ replaced by $l^*=n^*-1$, according to the ADK version of the PPT formula \cite{ammosov_jetp1986}. 
While we do not see justification for this step as well, we demonstrate in Section~\ref{sec:v} and Appendix~\ref{app:cnl}, that it does not significantly affect the precision of calculations. 

\subsection{Rate equations}
\label{sec:ivb}
Within this approach one has to solve a system of rate equations:
\begin{equation}\label{rateeq}
\frac{dn_{C_Z}}{dt}=\sum\limits_{C'}\left[n_{C'_{Z-1}}w_{C'_{Z-1}\to C_{Z}}- n_{C_Z}w_{C_Z\to C'_{Z+1}}\right],
\end{equation}
where $0\leq Z\leq Z_{\rm m}$, $n_{C_Z}$ is the concentration of ions with charge $Z$ in a state with electronic configuration $C_Z$ (e.g. $C_{14}=1s^22s2p$ is one of possible configurations for ${\rm Ar}^{14+}$) and $w_{C'_{Z'}\to C_{Z}}$ corresponds to the full transition probability rate from configuration $C'_{Z'}$ to $C_{Z}$. The summation over $C'$ accounts for possible ionization pathways for a given set of atomic configurations $\{C_{Z}\}$. 
Nonsequential ionization pathways can be included straightforwardly, and the dependence on the quantum numbers of each level can be rigorously accounted for. 

The rate equations approach has certain limitations. 
Firstly, the number of equations in  Eq.~(\ref{rateeq}) is equal to the number of possible intermediate configurations in the multiple ionization process, which quickly grows with $Z_{\rm m}$, and can reach several dozen for heavy atoms. 
Secondly, the simplest formulation Eq.~(\ref{rateeq}) neglects the ion motion as well as the effect of the self-consistent plasma field. 
Finally, although other ionization mechanisms (e.g., collisions) potentially can be included in the system Eq.~(\ref{rateeq}) after its significant modification (see examples in \cite{helios_06,pikuz_06} and references therein), they are typically disregarded. Notably, for PIC codes, such inclusion can be made in a straightforward way.

\subsection{Sequential vs nonsequential ionization pathway}
\label{sec:ivc}
\begin{figure}
    \centering
    \includegraphics[width=\linewidth]{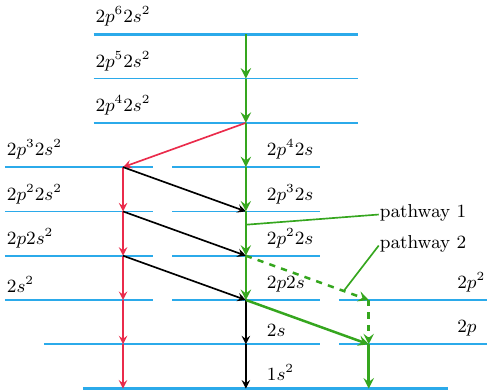}
    \caption{Major ionization pathways for multiple ionization of ${\rm Ar}^{8+}\rightarrow{\rm Ar}^{16+}$. Green solid and dashed arrows correspond to the dominant pathways, red arrows --- to the sequential one, black arrows --- to other possible pathways.}
    \label{fig:pathway}
\end{figure}

Let us consider the impact of sequential ionization approximation. 
We focus on the example of ${\rm Ar}^{8+}$ being ionized to ${\rm Ar}^{16+}$ by a short laser pulse, however, the scheme can be easily generalized to other ions. We assume that both the initial and final argon ions are in the ground state, i.e. their electronic configurations read $1s^22s^22p^6$ and $1s^2$, respectively. 
The final state $1s^2$ can be reached in multiple ways with sequential ionization being one of them. 
In Fig.~\ref{fig:pathway}, we show 12 sample ionization pathways relevant to the considered example.\footnote{We omit other possible pathways for transitions within the $2p-2s$ subshells for the sake of clarity. Our analysis shows that their contribution is significantly suppressed as compared to the selected 12 pathways.}

\begin{table*}
\caption{Ionization potentials (in atomic units), magnetic quantum numbers, and residual number of electrons with the same $|m|$ on a sublevel for three pathways, corresponding to the red, solid green, and dashed green arrows in Fig.~\ref{fig:pathway}. Values for $I_p$ are calculated using tables for the agron electronic energy levels from Ref.~\cite{saloman-argon}.}
\label{tab:Ip_pathways}
\begin{tabularx}{\linewidth}{l|lXXX|lXXX|lXXX} 
\hline\hline
$Z$ & State & $I_p$ & $m$ & $g_{|m|}$ &  State & $I_p$ & $m$ & $g_{|m|}$ &  State & $I_p$ & $m$ & $g_{|m|}$ \\ \hline
8 & $2p^6$ & 15.53 & 0 & 2 & $2p^6$ & 15.53 & 0 & 2 & $2p^6$ & 15.53 & 0 & 2 \\
9 & $2p^5$ & 17.633 & 0 & 1 & $2p^5$ & 17.631 & 0 & 1 & $2p^5$ & 17.631 & 0 & 1 \\
10 & $2p^4$ & 19.860 & 1 & 4 & $\boldsymbol{2s^2}$ & 21.892 & 0 & 2 & $\boldsymbol{2s^2}$ & 21.892 & 0 & 2 \\
11 & $2p^3$ & 22.745 & 1 & 3 & $2p^4$ & 20.713 & 1 & 4 & $2p^4$ & 20.713 & 1 & 4 \\
12 & $2p^2$ & 25.190 & -1 & 2 & $2p^3$ & 24.160 & 1 & 3 & $2p^3$ & 24.160 & 1 & 3 \\
13 & $2p$ & 27.750 & -1 & 1 & $2p^2$ & 26.700 & -1 & 2 & $\boldsymbol{2s}$ & 30.505 & 0 & 1\\
14 & $2s^2$ & 31.436 & 0 & 2 & $\boldsymbol{2s}$ & 32.608 & 0 & 1 & $2p^2$ & 28.680 & -1 & 2 \\
15 & $2s$ & 33.746 & 0 & 1 & $2p$ & 32.576 & -1 & 1 & $2p$ & 32.576 & -1 & 1 \\
\hline\hline
\end{tabularx}
\end{table*}	

\begin{figure}
    \centering
    \includegraphics[width=\linewidth]{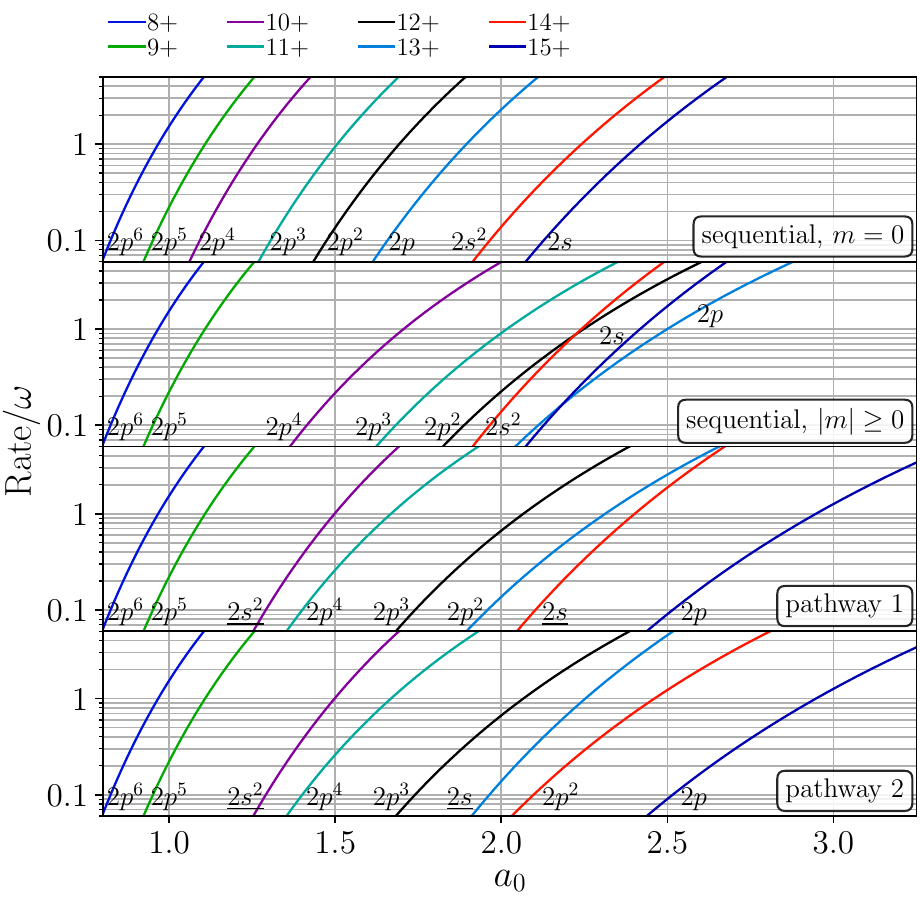}
    \caption{Dependence of the ionization rates [see Eq.~\eqref{PPT_definition}] on the dimensionless field amplitude $a_0$ for ${\rm Ar}^{Z+}$, $Z=8..15$, along three different ionization pathways. For all curves in the top plot, $m$ is set to zero, as in Ref.~\cite{nuter_pop2011}. The rest three plots are calculated using the data from Tab.~\ref{tab:Ip_pathways}. Here, $\omega$ is the laser frequency corresponding to the wavelength $\lambda=0.8$ $\mu$m.}
    \label{fig:rates_a0}
\end{figure}

In sequential ionization, the corresponding pathway goes through the ground states of the ions. 
In this case, the ionization potentials $I_p$ for the outermost electron are known for all charged states of most elements \cite{NIST}. For nonsequential ionization, the $I_p$ values are generally not directly accessible but can be calculated as follows. 
If an inner electron is extracted first, the residual ion appears an excited state. 
Energetically, the same state could be achieved by first ionizing the outermost electron and then exciting the residual ion. 
The excitation energies for different states of argon can be found in Ref.~\cite{saloman-argon}. 
To illustrate the procedure, consider extraction of the $2s^2$ electron from configuration $2s^22p^4$, which leads to the $2s2p^4$ state. Then, the ionization potential
\[I_p(2s^22p^4 \rightarrow 2s2p^4)=I_p(2s^22p^4 \rightarrow 2s^22p^3)+\Delta\varepsilon,\]
where the $\Delta\varepsilon\approx55.3$ eV is the excitation energy from $2s^22p^3$ to $2s2p^4$, and $I_p(2s^22p^4 \rightarrow 2s^22p^3)\approx540.4$ eV is the ground state ionization potential. 
We provide $I_p$ values for the three pathways in Tab.~\ref{tab:Ip_pathways}. 
Note that we neglect the relaxation of the excited states, as it happens on a larger time scale (nano and pico-seconds) than the laser-ion interaction time considered in this work (fs).

Let us study how the ionization rates given by Eq.~\eqref{PPT_definition} vary for different ionization pathways. Fig.~\ref{fig:rates_a0} shows the field dependence of the rates for the charge states ${\rm Ar}^{8+}..{\rm Ar}^{15+}$ for three sample pathways. Firstly, consider the case of sequential ionization. If we set $m=0$ following the hypothesis of Nuter et al. \cite{nuter_pop2011}, the rate curves corresponding to different charge states are ordered and separated (see the first subplot). 
That is, for given $a_0$, the probability to ionize a higher charged state is always significantly lower. 
Hence, under the $m=0$ approximation, ionization can indeed be viewed as sequential. 

Now, following our discussion in Section~\ref{sec:iiic2}, let us consider that the $m$ number is conserved during sequential ionization (see the plot labeled ``sequential, $|m|\geq0$'' Fig.~\ref{fig:rates_a0}). 
In this case, the ionization rate of the $2p^4$..$2p$ electrons (10+..13+) with $|m|=1$ is highly suppressed. Moreover, the rate for the $2p$ state (13+) appeared to be lower than for $2s^2$ and $2s$ electrons (15+, 16+) for the same $a_0$. 
This suggests that the sequential ionization pathway is not dominant.

After considering different nonsequential ionization pathways, we identified two (shown in green arrows in Fig.~\ref{fig:pathway}), for which the regular order of the ionization rate curves is restored, see plots labeled ``pathway 1'' and ``pathway 2'' in Fig.~\ref{fig:rates_a0}. One can note that the ionization rates for ${\rm Ar}^{10+}$..${\rm Ar}^{13+}$ in these cases are higher than the respective rates for sequential ionization (compare to the second subplot). Therefore, these nonsequential pathways contribute more than the sequential one. 

The two identified nonsequential pathways appear to be dominant in the considered example. This can be substantiated by comparing the solutions to the system of rate equations \eqref{rateeq} for all possible pathways and the dominant ones, which we do in Section~\ref{sec:vb}. Summarizing this subsection, multiple ionization is essentially nonsequential for close atomic sublevels, if the $m$ number can take nonzero values and is conserved.

\subsection{Dependence on the magnetic quantum number and nonsequential ionization pathways in PIC simulations}
\label{sec:ivd}
At the beginning of Section~\ref{sec:ii}, we listed the key assumptions limiting the inclusion of strong-field ionization in modern PIC codes. In view of our discussion in the previous subsection, the attempt to account for the dependence on quantum numbers correctly can contradict the requirement to treat ionization as sequentially proceeding along the pathway with progressively growing ionization potentials. 
However, upon a closer study, one may find that practical implementations like Ref.~\cite{nuter_pop2011} are reliant not on the actual sequence of ionization; rather, they require having a unique pathway of ionization. In effect, condition II introduced Section~\ref{sec:ii} can be reformulated in a more flexible way:
\begin{enumerate}
    \item[II.] For a given charge state, a level, outermost or inner, always exists, for which the ionization probability is dominant, and it is sufficient to account for only one dominant (sequential or nonsequential) ionization pathway.
\end{enumerate}  

We implement a refined numerical procedure to account for the dependence of ionization rates on $m$, assuming that the process is dominated by one pathway, which is not necessarily sequential. 
We assume that during the interaction with the field, $m$ is conserved for electrons residing on the atomic levels. Since sublevels sharing the same $|m|$ have equal ionization potentials (we assume that level splitting is negligible when calculating the rate), we consider that any of the corresponding electrons can be extracted. This can be taken into account by introducing the degeneracy factor $g_{|m|}$, namely, the number of residual electrons on the subshell multiplying Eq.~\eqref{PPT_definition}. Then the probability rate to extract an electron in a given configuration $C_Z$ reads:
\begin{equation}\label{rate_gm}
    w_{C_Z}=g_{|m|}w^{\rm PPT}[E(t); Z,I_p,l,m].
\end{equation}

Let us assume that an atom or ion is initially in the ground state. 
The numbers $m$ and $g_{|m|}$ are attributed to each electron as follows (see Tab.~\ref{tab:m_order} for an illustration): 
	\begin{enumerate}
		\item[1.] Electrons within a fixed $l$-subshell are ordered by increasing the value of $|m|$ (namely, the outermost electrons on the have $m=0$).\footnote{Within this algorithm, on a subshell with $l\geq1$ that is initially not full, states with lower $m$ appear first. For example, if there are three $p$ electrons, they will be assigned $m=0,\, 0, -1$. As an alternative, such electrons could be assigned a random value of $m=-l,\ldots,l$. Such a modification could be studied elsewhere.}
        
        \item[2.] The factor $g_{|m|}$ is the number of electrons left on a subshell with a given value of $|m|$, see Table \ref{tab:m_order} below.
        
	   \item[3.] The order of electron extraction is set by the dominant ionization pathway.
	\end{enumerate}
The dominant pathway for a given initial electronic configuration of an atom/ion should be identified individually, for example, by analyzing the rate dependence on the field strength (as in Section~\ref{sec:ivc}), or by solving the system of rate equations for different pathways. This can be done for the simplified case of a spatially homogeneous field. Once this pathway is identified, it can be used in the full 3D PIC simulation.

\begin{table}[h!]
        \caption{The order for electron sequential extraction from a sublevel with fixed $l$ at varying magnetic quantum number $|m|\leq l$ for ionization algorithm in PIC simulations.}
        \label{tab:m_order}
			\begin{tabularx}{\linewidth}{l XXXXXXXXXXX}
                \hline\hline
				Sublevel order & 1 & 2 &  3 &  4 & 5 & 6 &  7 &  8 & 9 & 10 & $\ldots$ \\
				$m$ & 0 & 0 & -1 & -1 & 1 & 1 & -2 & -2 & 2 & 2 & $\ldots$
                \\ 
				$g_{|m|}$ & 2 & 1 & 4 & 3 & 2 & 1 & 4 & 3 & 2 & 1 & $\ldots$
                \\\hline\hline
			\end{tabularx}
\end{table}

The proposed scheme allows us to generalize the algorithm proposed by Nuter et al \cite{nuter_pop2011} simply by replacing the ionization rates there without affecting the code performance. 
We developed a corresponding new ionization module for the PIC code SMILEI. Our implementation is available online \cite{github}. 

Note that the bottleneck of the proposed method is in the identification of the dominant pathway. 
Generally, it is not guaranteed that a single pathway or a small number of pathways will make the dominant contribution to ionization of higher-order subshells such as $d$ or $f$. However, we argue that using even one of such dominant pathways in a PIC implementation increases the precision of simulations. 
In our example for argon, calculations show that a couple of dominating ionization pathways give comparable contributions. As we show below, taking into account only one of them provides a reasonably accurate approximation.

\section{Numerical example: ionization of Argon}
\label{sec:v}
\begin{figure}
    \centering
    \includegraphics[width=\linewidth]{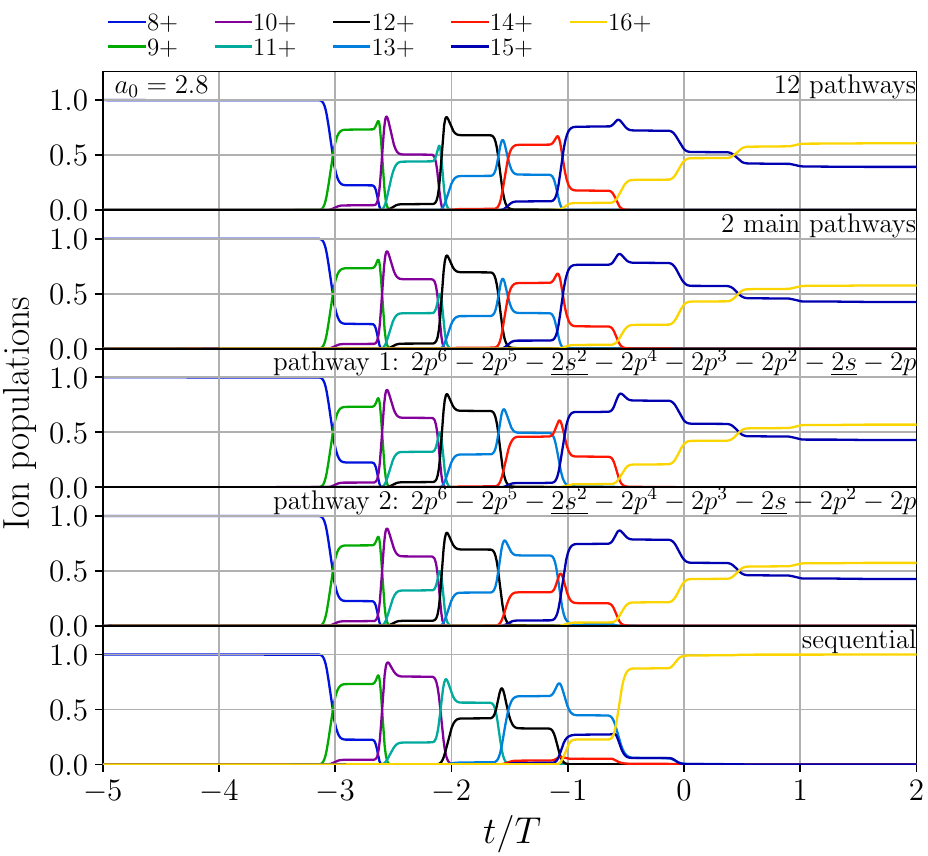}
    \caption{Time evolution of the ion populations $n_Z/n_0$ in a thin argon target prepared in the state ${\rm Ar}^{8+}$ (so that initially $n_8/n_0=1~,~n_Z=0~$ for $Z>8$) irradiated by a 10-cycle laser pulse. The curves are obtained by solving the rate equations \eqref{rateeq} in 1D for a different choice of pathways, see also Fig.~\ref{fig:pathway}. Time is normalized to the laser period $T$, the pulse front edge and its maximum arrive at the target front surface at $t=-5T$ and $t=0$, respectively. }
    \label{fig:rate_eqs_dynamics}
\end{figure}
\begin{figure}
    \centering
    \includegraphics[width=\linewidth]{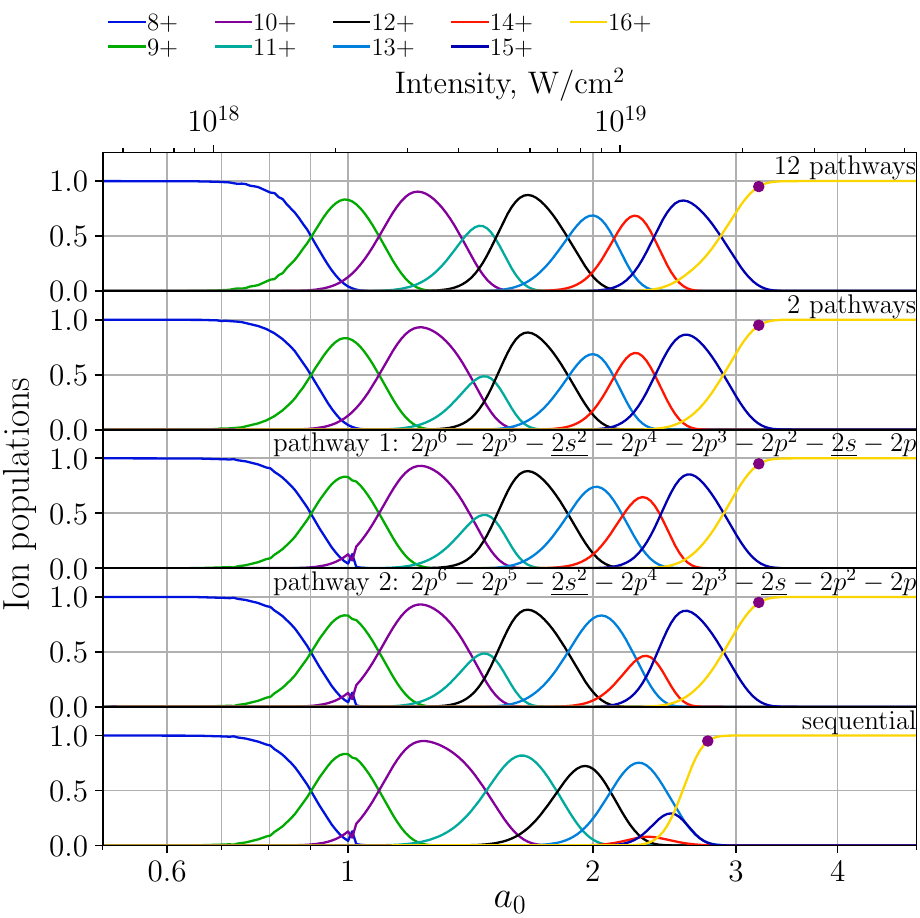}
    \caption{Residual ion populations left after the interaction as a function of the laser peak field strength (in the dimensionless units of $a_0$; also shown in the units of intensity in the top horizontal axis) obtained by solving the rate equations \eqref{rateeq} in one dimension for a different choice of pathways. Round points correspond to saturation of Ar$^{16+}$, when its population reaches 95\%.}
    \label{fig:rate_eqs_profiles}
\end{figure}

We consider the ionization of argon as an example that illustrates the statements and suggestions made in the previous sections. 
Argon, along with other noble gases, was widely used as a target starting from the early days of high-field physics \cite{huillier_pra89}.
For our considerations, argon is particularly suitable because its highest ionization potentials correspond to the ionization saturation intensity $\simeq 10^{21}{\rm W/cm}^2$, and the $2p2s$ shell is ionized at $\simeq 10^{18}-10^{20}{\rm W/cm}^2$ so that the ionization dynamics of interest happens at the highest intensities currently available at novel petawatt laser facilities.

We test the above-discussed schemes of treating tunneling ionization in simulation codes. 
First, we numerically solve the system of rate equations \eqref{rateeq} for various pathways and compare the outputs to identify the dominant one. 
Then we run PIC simulations with several ionization models to quantify the effect of the approximations discussed in Section~\ref{sec:iii}.

\subsection{Simulation setup}
\label{sec:va}
In all cases, we consider the interaction of a short intense laser pulse with a thin ($\ll \lambda$) layer of gaseous argon\footnote{The narrow target is chosen in order to boost the speed of the simulations. We checked that filling the entire simulation box with argon does not change the results.}. 
The laser field with the carrier frequency $\omega$ corresponding to the wavelength $\lambda=0.8$ $\mu$m is linearly polarized along the $y$ and propagates along the $x$-axis: $\vec{E}= a_0f(t-x/c)\{ 0, \cos(\omega t-kx),\, 0\}$. The pulse envelope has the shape:
\begin{equation}
    f(\varphi)=\left\lbrace 
        \begin{array}{ll}
         \cos^2\left( \dfrac{\varphi}{2 N_{\rm cyc}}\right), &  -N_{\rm cyc}\pi  \leq \varphi < N_{\rm cyc}\pi ,\\
         0, & \text{otherwise}.
    \end{array}
    \right.
\end{equation}
where $\varphi=\omega t -kx$, $k=2\pi/\lambda$ and $N_{\rm cyc}=10$ is the total number of cycles in the pulse. 
The phase is chosen such that the field reaches its absolute maximum at the maximum of the envelope, $\varphi=0$. 
The field amplitude $a_0$ is varied in simulation runs. 

Initially, argon (neutral or precharged) experiences zero field. 
As a simulation starts, the laser pulse impinges on the target and ionizes it as the pulse propagates through. 
We set $t=0$ as the time at which the maximum of the laser envelope reaches the target front surface. 
The laser pulse starts interacting with argon at $t=-5T$, and most of ionization events occur at $t<0$.  
We keep the initial argon density $n_0=5\times 10^{13}$ cm$^{-3}$ sufficiently low so that collective effects are negligible. 
In our analysis, we focus on relative ion populations $n_Z/n_0$, $0\le Z\le Z_{\rm m}$. 

For PIC simulations, we use a customized version of the PIC code SMILEI. \footnote{At the moment when this publication has been submitted, the official version of SMILEI \cite{derouillat_cpc2018} follows Ref.~\cite{nuter_pop2011} relying on the sequential ionization hypothesis and using the ADK formula with $m$ set to zero. 
The option to account for the full dependence on $m$ and and $g_{|m|}$ via the PPT formula \eqref{PPT_definition} was also recently included as an experimental feature. 
TL and KAG models were also added.}  
It allows us to define the rates via the PPT formula \eqref{PPT_definition} with different prescriptions for evaluating $C_{n^*l}$, account for the dependence on $m$ and $g_{|m|}$ under the approximation of conserved quantum numbers (see Section~\ref{sec:ivd}), include a (single) full custom ionization pathway (with the ionization potentials corrected as suggested in Section~\ref{sec:ivc}), and the possibility to combine all of these features with the TL and KAG models for the BSI corrections. Our new implementation of SMILEI, as well as sample input files used in this work, are available on GitHub \cite{github}.

We simulate the described setup in one dimension. The simulation box total length is $8\lambda$, and the argon target occupies the 4 central cells. 
At the resolution 128 points per wavelength, this amounts to $\approx 0.03\lambda$. 
Such resolution was chosen as a result of convergence checks to ensure the simulation precision. 
We also choose 8192 particles per cell to ensure high-quality statistics of the Monte-Carlo simulation of the ionization process.

\subsection{Dominant ionization pathway from the rate equation solution}
\label{sec:vb}
Let us study our setup using rate equations \eqref{rateeq}. We assume that argon is initially charged to Ar$^{8+}$ (with the electronic configuration $2p^62s^2$) and focus on the transition to Ar$^{16+}$ ($1s^2$ final state). We solve the system numerically for different pathways shown in Fig.~\ref{fig:pathway} by taking into account: (i) all 12 pathways, (ii) two main pathways (summed up and considered each separately) labeled ``pathway 1'' and ``pathway 2'', and (iii) the sequential pathway. 
We use Eq.~\eqref{rate_gm} for the transition probability rates in Eqs.~\eqref{rateeq} and account for the dependence on the $m$ number as discussed in Section~\ref{sec:ivd}. The full system is lengthy for presenting here (e.g., it includes 15 equations for 12 pathways); however, it can be straightforwardly obtained with the aid of Fig.~\ref{fig:pathway}. The results are presented in Figs.~\ref{fig:rate_eqs_dynamics} and \ref{fig:rate_eqs_profiles}.

Fig.~\ref{fig:rate_eqs_dynamics} shows time-dependent ionization dynamics for a particular case of $a_0=2.8$.
This amplitude is close to the ionization threshold of Ar$^{15+}$. 
The solutions for all twelve and only two main pathways (shown in the first and second plots of Fig.~\ref{fig:rate_eqs_dynamics}, respectively) are close to each other for all intermediate and final ion states. This confirms our guess (see Section~\ref{sec:ivc}) that the two green pathways in Fig.~\ref{fig:pathway} are dominant. 
The solutions accounting for only one of these pathways, shown in plots \#3 and 4 in Fig.~\ref{fig:rate_eqs_dynamics}, are also visually close to the full 12-pathway solution.

The case of the sequential ionization (shown in the bottom plot of Fig.~\ref{fig:rate_eqs_dynamics}) is noticeably different. Firstly, ionization of Ar$^{12+}$ and Ar$^{13+}$ (black and light-blue curves, respectively) occurs later, i.e. when a higher field arrives at the target. This is consistent with Fig.~\ref{fig:rates_a0}: the probability rates for ionizing Ar$^{12+}$ and Ar$^{13+}$ in sequential order are lower than those of pathways 1 and 2.
Secondly, Ar$^{14+}$ and Ar$^{15+}$ barely appear during ionization in the sequential order (red and dark-blue curves, respectively, in the bottom plot of Fig.~\ref{fig:rate_eqs_dynamics}). The field needed to ionize Ar$^{13+}$ exceeds the threshold of Ar$^{15+}$ ionization (second plot in Fig.~\ref{fig:rates_a0}), hence, as it is reached, the ion rapidly transforms into Ar$^{16+}$. 
These features are not seen in the full solution for 12 pathways (which includes the sequential one). 
Finally, the residual ion population after the interaction is also noticeably different for the sequential order.

To check the sensitivity of these results to the laser peak intensity, we run calculations for different amplitudes $a_0$. Figure~\ref{fig:rate_eqs_profiles} shows the residual ion populations (from now on referred to as the ion profiles) as a function of the peak field strength. 
As before, while the reduced system containing pathways 1 and 2 gives a solution close to the full 12-pathway case, the sequential order approximation predicts a noticeably different outcome. 

For instance, the threshold intensity needed to produce Ar$^{16+}$ ions changes. 
Here we define the saturation field strength for a given ionic state as that at which the ion population reaches 95\%.\footnote{Note that such a definition is only consistent for ionic states separated from the next shell by a wide gap in the ionization potential as it takes place for Ar$^{16+}$.
For states inside a given $n$-shell the 95\% may never be achieved because of subsequent ionization steps follow quickly (see also Fig.~\ref{fig:error_saturation} below and the discussion around).}
Then for Ar$^{16+}$, saturation is reached at $a_0=2.77$ for sequential ionization and at $a_0=3.20$ in the 12-pathway case (shown with points in the corresponding plots in Fig.~\ref{fig:rate_eqs_profiles}). The former underestimates the saturation field by 13\%. The error for Ar$^{14+}$ and Ar$^{15+}$ states is higher and even exceeds 50\% at maximum. For more details on the absolute error see Fig.~\ref{fig:errors_rate_eqs} in Appendix~\ref{app:errors}.
Here, one should note that in Fig.~\ref{fig:rate_eqs_profiles}, the sequential ionization pathway looks ``faster'' and therefore ``more efficient'' than the others.
However, when all possible (including the two most relevant) ionization pathways are accounted for, quick ionization of the $2s^2$ electron blocks the further development of the sequential channel and eventually shifts the saturation threshold towards higher intensities.

From these results, we conclude that using only pathway 1 or 2 provides a reliable approximation to the full 12-pathway solution. The absolute error for most curves shown in plots 3 and 4 of Fig.~\ref{fig:rate_eqs_profiles} is below 0.1 at maximum. We find that \textit{pathway 1} provides a higher precision result in a wide range of $a_0$ (see also Fig.~\ref{fig:errors_rate_eqs} in Appendix~\ref{app:errors}), therefore, we choose it as the dominant pathway describing the ionization of the Ar$^{8+}(2p^62s^2)$ shell. 
We use it in PIC simulations as suggested in Section~\ref{sec:ivd}.

\subsection{Implementation in the PIC code}
\label{sec:vc}
\begin{figure}
    \centering
    \includegraphics[width=\linewidth]{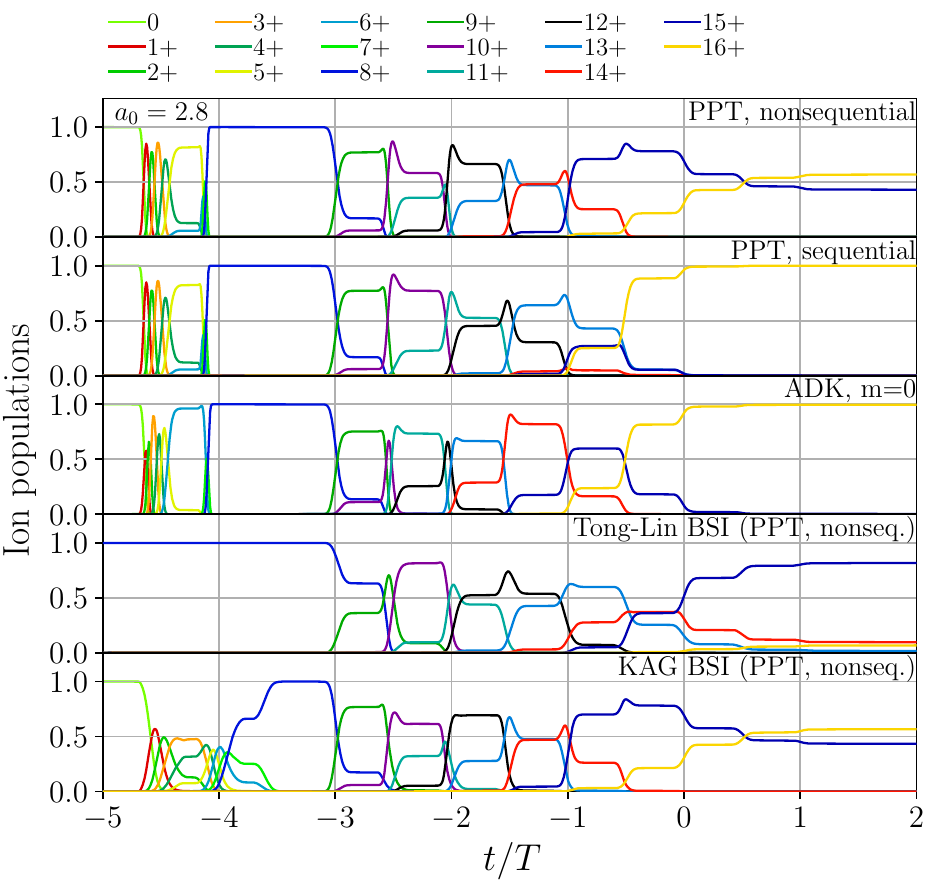}
    \caption{Evolution of ion populations $n_Z/n_0$ extracted from 1D PIC simulations of a thin argon target irradiated by a 10-cycle laser pulse. In all cases except plot 4, the initial state is neutral argon, while the results of plot 4 are for argon pre-ionized up to the ${\rm Ar}^{8+}$ state. Plots show the results obtained within different approximations (see the legends).}
    \label{fig:PIC_ionization_dynamics}
\end{figure}

Now let us consider a PIC realization of the described setup. Our main goal is to identify the impact of different approximations on the simulation accuracy. 
Following the discussion of potential error sources in Section~\ref{sec:iii}, we test the effects of: (i) accounting for $|m|\geq 0$ (following Section~\ref{sec:ivd}) or using the approximation $m=0$ (following \cite{nuter_pop2011}), (ii) the impact of nonsequential ionization, (iii) barrier suppression modifications of the ionization rate within the TL and KAG models, (iv) choice $C_{n^*l}$ in Eq.~\eqref{PPT_definition}. Results are presented in Figs.~\ref{fig:PIC_ionization_dynamics} and \ref{fig:PIC_profiles} showing ionization dynamics and ion profiles, respectively, and also in Appendix \ref{app:cnl}.

\begin{figure}
    \centering
    \includegraphics[width=\linewidth]{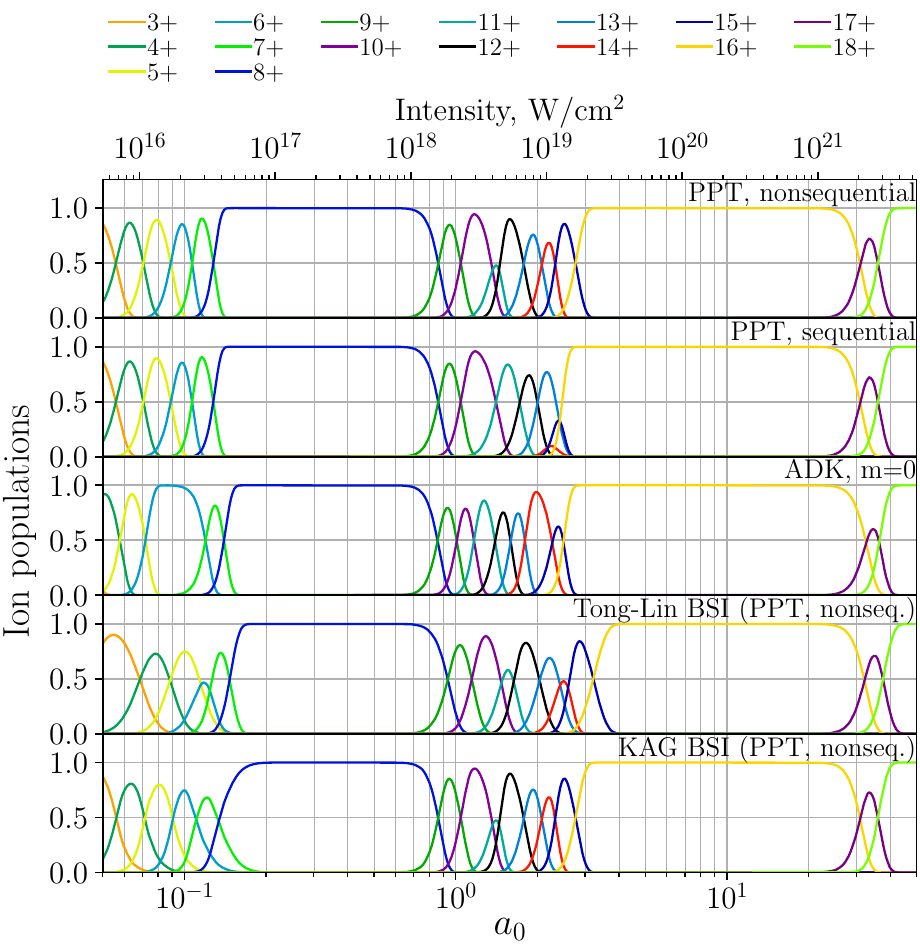}
    \caption{Residual ion populations after the interaction as functions of the laser pulse peak field in units of $a_0$ (bottom horizontal axis) and of the intensity (top horizontal axis) obtained from 1D PIC simulations. Notations are the same as in Fig.~\ref{fig:PIC_ionization_dynamics}.}
    \label{fig:PIC_profiles}
\end{figure}
\begin{figure}
    \centering
    \includegraphics[width=\linewidth]{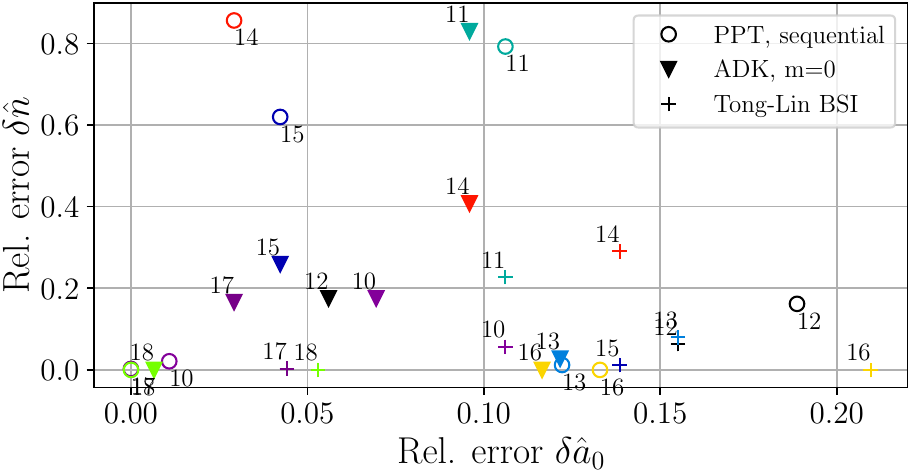}
    \caption{Relative error of an ion population maximum location in $\delta \hat{a}_0$ [see Eq.~\eqref{error_def}] and value $\delta \hat{n}$ for $Z\geq 10$ (for $Z=16$ and $18$, the saturation point $n_Z/n_0=0.95$ is taken). The error is calculated from Fig.~\ref{fig:PIC_profiles} using Eq.~\eqref{error_def}, the models denoted in the legend are benchmarked against ``PPT, nonsequential''. A number near each marker shows the ion charge state.
    }
    \label{fig:error_saturation}
\end{figure}

In PIC simulations, we start from a neutral target (except for the TL BSI model for the reason explained below). 
In all cases, the first eight electrons populating the $3p$ and $3s$ subshells, are stripped off quickly, see Fig.~\ref{fig:PIC_ionization_dynamics}. Their ionization thresholds are below $a_0\sim 0.1$, which is much lower than the threshold for Ar$^{8+}$, $a_0\sim 1$, see Fig.~\ref{fig:PIC_profiles}. 
As a result, ionization dynamics of these outer shells does not affect the distributions of ions with $Z\geq 8$. 
This is also confirmed by comparing two simulations: one with an initially neutral target, and the other prepared in the charge state $Z = 8$, which provide the same results for $Z>8$ ions. 
In effect, as long as deep ionization states are considered, it is enough to study the strong-field ionization model modifications only for the inner subshells. 
For simplicity, we treat the ionization of the outer states $Z=0..7$ as sequential proceeding according to the PPT rates and focus on studying the $2p$, $2s$, and $1s$ subshells.

\subsubsection{Nonsequential ionization and $|m|\geq0$}
\label{sec:vc1}
First, we confirm that PIC simulations provide results identical to those obtained from the rate equation solution. The top plots in Figs.~\ref{fig:PIC_ionization_dynamics} and \ref{fig:PIC_profiles} labelled ``PPT, nonsequential'' show the simulation results obtained with our improved module for ionization in the PIC loop: the rates are calculated via Eq.~\eqref{PPT_definition} with $C_{n^*l}$ given by the Hartree formula \eqref{hartree}, accounting for the full dependence on $|m|\geq0$ and the dominant nonsequential ionization \textit{pathway 1}. The results for the states with $Z\geq 8$ perfectly match the data obtained by solving the rate equation (compare with the plots labeled ``pathway 1'' in Figs.~\ref{fig:rate_eqs_dynamics} and \ref{fig:rate_eqs_profiles}). The simulation results for sequential ionization obtained with the two methods also match perfectly.

Next, we compare the approach of Nuter et al \cite{nuter_pop2011} with the case when the $m$ number dependence is included, see the corresponding plots labeled ``ADK, $m=0$'' and ``PPT, sequential'' in Figs.~\ref{fig:PIC_ionization_dynamics} and \ref{fig:PIC_profiles}. 
When compared to the full model (``PPT, nonsequential''), a noticeable difference is seen for $Z=8..16$, which shows that the choice of model for treating the magnetic quantum number is important.
At the same time, the ion profiles for $Z=17$ and $18$ in Fig.~\ref{fig:PIC_profiles} are almost independent of the model choice, which is not surprising taking into account a wide gap in the ionization potential between the $2s$ and $1s$ shells. 

To highlight the difference between the models, let us consider the positions of the ion profile maxima for states $Z\geq 10$ (see Fig.~\ref{fig:PIC_profiles}). 
For $Z=16$ and $18$, which reach a clear saturation, we look at the points $n_{Z}/n_0=0.95$ instead of the maximum. 
For the model labeled ``x'' and for each $Z$ profile, we denote the corresponding point location and value by $a_0=\hat{a}_{0,\,{\rm x}}(Z)$ and $n_Z=\hat{n}_{Z,\,{\rm x}}$, respectively. 
As seen in the figure, for fixed $Z$, $\hat{a}_{0,\,{\rm x}}(Z)$ and $\hat{n}_{Z,\,{\rm x}}$ vary with the choice of the model. 
We quantify the deviation from our primary model (``PPT, nonsequential'') by the relative errors $\delta \hat{a}_0$ and $\delta \hat{n}$ for each $Z$. The former reads:
\begin{equation}
    \label{error_def}
    \delta \hat{a}_0=\frac{|\hat{a}_{0,\,{\rm x}}-\hat{a}_{0,\,{\rm PPT, nonseq}}|}{\hat{a}_{0,\,{\rm PPT, nonseq}}}.
\end{equation}
Analogously, we define $\delta \hat{n}$. 

In effect, the deviation of a maximum position can be represented by a point in coordinates $(\delta \hat{a}_0,\,\delta \hat{n})$, as shown in Fig.~\ref{fig:error_saturation}. As follows from the figure,  models ``PPT, sequential'' and ``ADK, $m=0$'' provide the same level of precision overall for the $2p$ and $2s$ states. The $\delta \hat{a}_0$ error stays at $\sim 0.1$ for most states. The error $\delta \hat{n}$ depends strongly on the state considered. 
The model ``PPT, sequential'' gives a slightly better prediction for the population values, as $\delta \hat{n}$ is generally lower than in the case of ``ADK, $m=0$'' (except for the states $Z=14$ and $15$ for the reasons discussed in Section~\ref{sec:vb}). 

From this analysis, we conclude that the main limitation for the precision of models ``PPT, sequential'' and ``ADK, $m=0$'' stems from the sequential ionization approximation. 
When the dependence on the $m$ number is considered, accounting for the dominant nonsequential pathway of ionization is crucial for improvement of the calculation accuracy. 

\subsubsection{Barrier suppression ionization}
\label{sec:vc2}
\begin{figure}
    \centering
    \includegraphics[width=\linewidth]{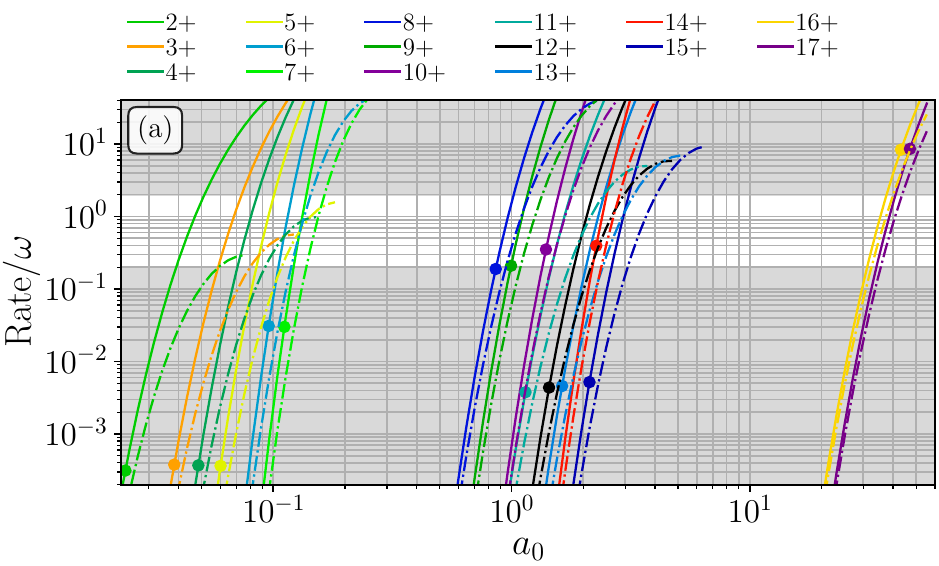}\\
    \includegraphics[width=\linewidth]{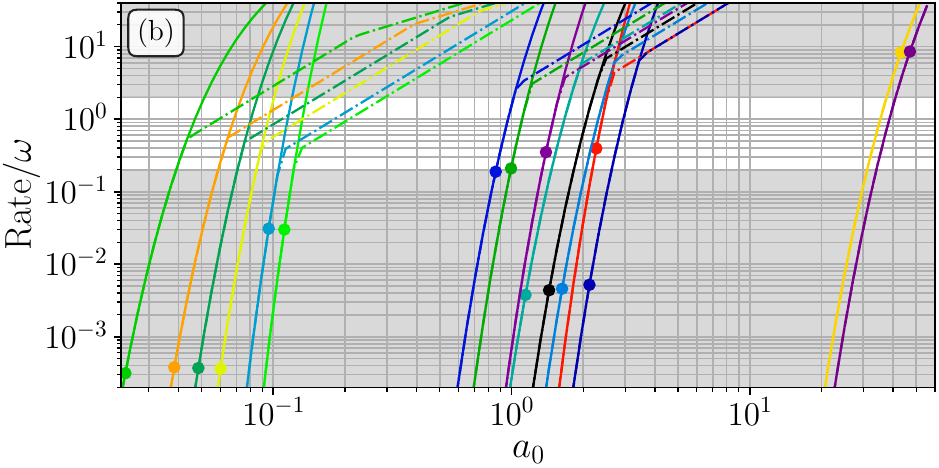}
    \caption{Dependence of the ionization rates on the field strength. In both plots, solid lines show the PPT formula \eqref{PPT_definition}. Round markers show the PPT rate value at the field $E_{\rm BS}$ [see Eq.~\eqref{BSIcond}]; at $E\gtrsim E_{\rm BS}$ the PPT rates require a BSI correction. Dot-dashed lines show the rates that include such a correction within the models of (a) Tong and Lin at $\alpha=6$ [see Eq.~\eqref{rate_TL}], (b) Kostyukov-Artemenko-Golovanov [see Eq.~\eqref{rates_KAG}]. Ion charge states are taken in the order of \textit{pathway 1} (see Fig.~\ref{fig:pathway}). In the interaction with a field of frequency $\omega$, the ionization event is most likely to happen in the unshaded area. }
    \label{fig:rates_BSI}
\end{figure}

Let us now consider the effect of BSI. 
Results of the two BSI models are shown in panels labeled ``Tong-Lin BSI'' and ``KAG BSI'' in Figs.~\ref{fig:PIC_ionization_dynamics} and \ref{fig:PIC_profiles}. In simulations, we take the full model ``PPT, nonsequential'' and modify it by using Eqs.~\eqref{rate_TL} and \eqref{rates_KAG}, respectively. We set the TL model parameter to $\alpha=6$.\footnote{We are not aware of empirical or numerical data that can be used to deduce $\alpha$ for argon BS ionization. We choose $\alpha=6$ following Refs.~\cite{tong_jphysb2005, ciappina_lpl2020} for the sake of an estimate. One may note (\ref{rate_TL}) that the true parameter of the TL model is $\alpha n^*/8\simeq 1$, which substantiates the choice of $\alpha$.} 

Recall that the TL correction works for fields $E\lesssim E_{\rm BS}$, where the latter is given by Eq.~\eqref{BSIcond}. We compare the field dependence of the PPT rates [Eq.~\eqref{PPT_definition}] and the TL-corrected rates [Eq.~\eqref{rate_TL}] in Fig.~\ref{fig:rates_BSI}a for different charge states of argon. In particular, we mark the locations of $E_{\rm BS}$ for each state. In the parameter range, where the ionization probability is significant, namely, $w/\omega\sim 1$ (see the unshaded region; $w$ is the ionization rate), ionization of the outer electrons (curves $Z\leq 5$) occurs deeply in the BSI regime $E>E_{\rm BS}$. 
The introduction of the TL factor leads to an unphysical underestimation of the rates.\footnote{When the TL formula is used in a simulation with an initially neutral target, strong-field ionization appears blocked due to this strong and physically unjustified suppression, leading to clearly erroneous results.} For deeper shells, $E\sim E_{\rm BS}$, which is close to the applicability range of the TL formula. To ensure the TL model robustness, it is advisable to run simulations with pre-ionized targets. In our case, we start from Ar$^{8+}$.

The main effect of the TL model on the ionization dynamic and profiles (see Figs.~\ref{fig:PIC_ionization_dynamics} and \ref{fig:PIC_profiles}) is the shift of the ionization thresholds towards higher intensities. 
As shown in Fig.~\ref{fig:error_saturation}, the ion population curves are systematically shifted in the horizontal direction by $\delta \hat{a}_0>0.1$ for $Z=10..15$, and $>0.2$ for the threshold of Ar$^{16+}$ saturation. 
At the same time, the ionization threshold for Ar$^{17+}$ and Ar$^{18+}$ are barely affected by the variations in the ionization model. 
The robustness of the $1s$ shell to the choice of the ionization model was analyzed and explained in  Refs.~\cite{ciappina_lpl2020,ciappina2020focal}.
This stability, which makes ionization of these states weakly sensitive to so far unresolved theoretical issues is particularly important for potential applications including laser intensity measurements \cite{ciappina_lpl2020}.

The KAG model extrapolates the rates to the range $E> E_{\rm BS}$. We plot the corresponding rates in Fig.~\ref{fig:rates_BSI}b. The main effect of the model is seen for outer electronic shells, while for the states $Z>8$ the model deviates from the PPT formula only when $w/\omega>1$. 
As we mentioned in Section~\ref{sec:iiib}, in an interaction with any reasonably smooth laser pulse, all ionization events will take place before the field $E>E_{\rm BS}$ is reached. 
As a result, our argon benchmarking example shows (see Figs.~\ref{fig:PIC_ionization_dynamics} and \ref{fig:PIC_profiles}) that while the dynamics of the outer states with $Z<8$ is sensitive to the model choice and the standard PPT is not always sufficient, deeper shells remain unaffected by the KAG modification of the ionization rate.

The accuracy of the BSI model is generally poor and can be improved only by the proper choice of the model constants made on the basis of comparisons with results of exact numerical solutions to the Schrodinger equation or with experimental data [see Eqs.~\eqref{rate_TL} and \eqref{rates_KAG}]. When such data are unavailable, the TL and KAG models can be used to test whether a simulation or a physical setup is sensitive to the BSI correction. 
The choice between the TL and KAG models should be made based on the analysis of the ionization rate curves, as we do for argon in Fig.~\ref{fig:rates_BSI}. 
As a general recommendation, before running a PIC simulation, it is advisable to check an overlap between the TL model and the intensity interval where we expect significant depletion of a given atomic level. 
If the process is likely to happen at $E>E_{\rm BS}$ (e.g. for outer electronic shells or if the field is switched abruptly), the KAG model can be applied, but it should be ensured that it provides a sensible impact on the considered ion states. Unfortunately, both models are not quantitatively predictive in the range $E\sim E_{\rm BS}$.

\section{Summary and discussion}
\label{sec:vi}
We discussed the treatment of strong-field ionization in PIC simulations in the context of laser-plasma interactions at high intensities. 
Although the tunneling ionization rates have been known for decades \cite{perelomov_jetp1966,perelomov_jetp1967, popov_ufn2004,poprz_jpb14}, their implementation in PIC codes may still meet serious difficulties.
The two most fundamental are: (a) inadequacy of the sequential picture of tunnel ionization and (b) breakdown of the semiclassical tunneling approximation in the high-field regime. Concerning issue (a), we demonstrate the necessity to determine the most probable pathway for ionization within $p$ subshells. We anticipate this is also relevant for subshells with higher $l>1$.
Issue (b) requires matching the PPT ionization rates with empirical formulas qualitatively describing the ionization probability in the BSI regime.

The insufficiency of the sequential ionization paradigm, assuming that electrons are removed from the ion in the order determined by the increase of their ionization potentials, is closely connected with the choice of model for treating the impact of the magnetic quantum number $m$.
Setting recursively $m=0$ for an outer electron \cite{nuter_pop2011} makes the sequential ionization pathway always dominant, which greatly simplifies the implementation in a PIC code.
Yet, for low-density gases in the intense field of a femtosecond laser pulse, we do not see arguments justifying this ansatz.
Assuming conservation of $m$, we demonstrated that the dependence of the rates on $m$ is of principal importance and makes \textit{nonsequential ionization pathways more probable than the sequential one}.
This in turn necessitates a significant reconsideration of the whole scheme.

Performing such a revision by taking the example of the $2p^62s^2$ shell in argon, we showed that a higher precision can be achieved by replacing the standard sequential order of ionization with the dominant nonsequential pathway. 
We implemented this approach in a customized strong-field ionization module in the SMILEI PIC code (available on GitHub \cite{github}). The nonsequential pathway has to be identified individually for each $n$-shell of each element by solving the system of rate equations.
Once the ordering of the most probable (sequential or nonsequential) ionization pathways is determined for a given atom, its implementation in a PIC code is straightforward.
A similar reconsideration of the ionization order can be made for heaver atoms, where $d$ and $f$ sub-shells will have to be included in the scheme.
For lighter atoms, such as neon or, especially, nitrogen, only several electrons are expected to be ionized in the single-electron tunneling regime, whereas the detachment of 3-4 outermost electrons will proceed in a specific correlated regime as described in Section II.

Note that our treatment still stays within the single-electron picture, and all the discussed models discard the possible impact of collective effects and electron-electron correlations in the ionization rate.
In high fields, single-electron treatment is well validated except for the case of extremely short laser pulses of few-fs and sub-fs duration, where correlated or anti-correlated tunneling of two or more electrons may become significant \cite{poprz_bull23,poprz_pr25}.

The impact of BSI deviations from the PPT rates is also found significant although less important than that of the nonsequential ionization.
Here, the main problem with development of improved schemes stems from the fact that no accurate analytic expressions for the rates exist in the BSI intensity domain.
The known models provide only qualitative description leading to unavoidable and poorly controllable losses in accuracy.
The main qualitative effect of BSI is clear: \textit{the ionization threshold shifts to higher intensities}. 
BSI modifications of the ionization rates appear more important for $l\ne 0$.
This, in combination with the dependence of the rates on the $m$ number leading to the competition between sequential and nonsequential ionization pathways, brings us to the conclusion that ionization of the inner $1s$ shell is described by the currently existing theory with the highest accuracy.

From the practical viewpoint, the scheme accounting for nonsequential ionization and BSI modifications demonstrated in this paper can improve PIC simulation precision up to one order of magnitude for certain ion states, see Fig.~\ref{fig:error_saturation}. 
The two most significant remaining sources of errors are: (i) the contribution of alternative ionization pathways and (ii) uncertainties (including empirical parameters) in the BSI models. 
The first can be controlled by cross-checking with the rate equations. 
For the second, one should rely on further experimental verification of the empirical formulas.

Finally, let us briefly cover possible applications of our results.
Methods to calculate the distributions in charge states $Z$ resulting from the ionization of heavy atoms in the focus of an intense laser pulse are of crucial importance for the realization of ionization-based techniques for \textit{in situ} diagnostics of ultra-high laser intensities \cite{ciappina_pra2019, ciappina_lpl2020, ouatu_pre2022}. 
The scheme demonstrated here for the case of argon shows that one may use ionization of different shells to control corresponding intensity intervals, while in each case the accuracy of diagnostics has to be quantified using the methods introduced in this paper.
To improve the reliability of the intensity diagnostics, one should observe the simultaneous ionization of two or more gaseous elements, say argon and xenon. 
In this case, ionization potentials may densely cover the probed interval of intensities, and the observation of simultaneous ionization of different electronic shells in different elements will help reduce the measurement uncertainties.

The presence and even dominance of nonsequential ionization pathways in the integrated ionization signal can be of interest for the production of excited ionic states through strong-field ionization.
For the example of argon considered here, ions  Ar$^{12+}$ with a $2s$ electron removed in the presence of $2p$ electrons demonstrate the presence of this ionization-excitation channel. 
Detection of photons emitted during the subsequent relaxation of these ``hole'' ions into their ground state can allow experimental indication of such specific ionic states.
\textit{Such an observation can also demonstrate the presence of the nonsequential ionization process} so far unobserved under high-field conditions.

High-field ionization of heavy atoms can also of interest as an efficient source of fast electrons created inside the focus.
Such ionization injector of electrons can be used for laser wake-field acceleration \cite{mcguffey_prl2010, pak_prl2010, pollock_prl2011, thaury_scirep2015}, for initiation of QED cascades \cite{artemenko_pra2017}, and for production of intense X-rays emitted during the acceleration of these electrons in the laser focus. Overall, our method will be of use for numerical exploration of various laser-plasma processes containing multiple ionization of atoms as an essential stage.

\section*{Acknowledgments}
This work was supported by Mobility plus project CNRS 23-12 from Czech Academy of Sciences and Centre national de la recherche scientifique, and by Sorbonne Universit\'e in the framework of the Initiative Physique des Infinis (IDEX SUPER). S.V.P. acknowledges support from the Russian Science Foundation (Grant No. 25-22-00308).

\appendix
\section{Comparison of the approximations for the asymptotic coefficients $C_{n^*l}$}
\label{app:cnl}
\begin{figure}
    \centering
    \includegraphics[width=\linewidth]{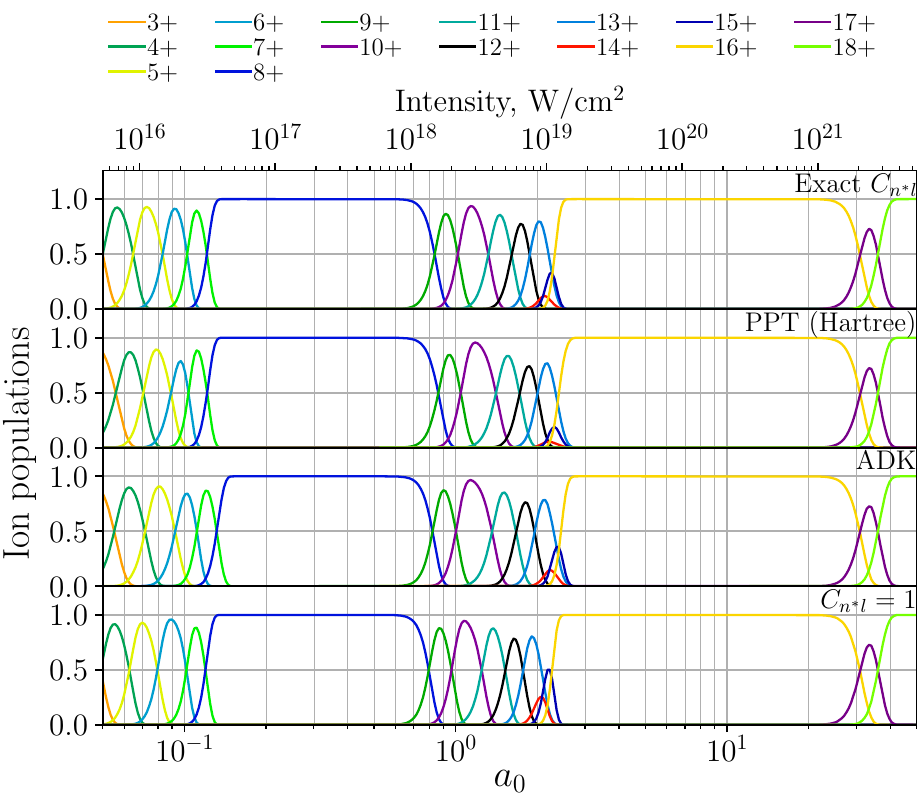}
    \caption{Ion profiles as a function of the laser pulse peak field strength in the dimensionless units of $a_0$ (also shown in the units of intensity in the top horizontal axis) obtained from 1D PIC simulations. In all simulations, we use the PPT formula \eqref{PPT_definition} for the ionization rates with different approximations for $C_{n^*l}$: (first) exact values extracted from the asymptotic numerical solution to the Schroedinger equation, (second) Hartree formula \eqref{hartree}, (third) ADK approximation [$l$ is replaced by $l^*=n^*-1$ in Eq.~\eqref{hartree}], and (fourth) $C_{n^*l}=1$ for all states. In all cases, we account for the dependence on $m$ as described in Section~\ref{sec:ivd}.}
    \label{fig:profiles_cnl}
\end{figure}

Here we examine the impact of different approximations for the asymptotic coefficients $C_{n^*l}$ in the tunneling formula \eqref{PPT_definition} on the value of the rate. 
Following the original approach \cite{perelomov_jetp1966, perelomov_jetp1967, perelomov_jetp1967b, popov_ufn2004}, $C_{n^*l}$ is a coefficient in the single-electron asymptotic ($r\to\infty$) of the radial wave function of a level with ionization potential $I_p$ and angular momentum $l$.
This coefficient can be exactly analytically expressed only for H-like ions. 
For many-electron atoms and ions it can be calculated in the Hartree approximation \eqref{hartree}. 
In the ADK formula, Hartree formula (\ref{hartree}) was modified by replacing $l$ with $l^*=n^*-1$ (although the dependence on $l$ in the $B_{l|m|}$ in Eq.~\eqref{PPT_definition} was kept). As mentioned in Section~\ref{sec:iiic1}, this can be a potential source of discrepancies, which we estimate here.

The coefficients $C_{n^*l}$ can be extracted from matching the numerical solution
\begin{equation}
    \psi^{\rm HF}_{nlm}(\mathbf{r})=\frac{P_{nl}(r)}{r}Y_{lm}\left(\frac{\mathbf{r}}{r}\right)
\end{equation}
of the Hartree-Fock (HF) integro-differential equations with an
exponentially decreasing solution (\ref{psi}) of the Schr\"odinger equation for the H-like
ion with the charge $Z$ at some point $r_m$ \cite{Clementi_1974,Evseev_1978,radzig_book2012,vitanov1991asymptotic}. 
The main problem here is the optimal choice of the matching point $r_m$. 
On the one hand, the value of $r_m$ must be large enough so that the radial HF function $P_{nl}(r)$ reaches the asymptotic with good accuracy, on the other hand, as $r_m$ increases, the accuracy of the HF function $P_{nl}(r)$ calculation drops.

The radial function $P_{\rm nl}(r)$ has been obtained
with a much higher accuracy then before \cite{Clementi_1974,Evseev_1978,radzig_book2012,vitanov1991asymptotic} by numerically solving the HF integro-differential equations using the finite difference
method. 
Therefore, we were able to use much higher values of the
matching point $r_{m}$ , precisely, where the electron density
$P^2_{\rm nl}(r)$ equals to 0.01 of its maximum value. 
We also verified that over a sufficiently wide range around the chosen $r_{\rm m}$ point, the constant $C_{n^*l}$ remains unchanged by 4-5 significant digits.

We also used the correction of the value of the asymptotic constant 
$A_{\rm HF}$ proposed in \cite{Evseev_1978} due to the fact 
that the energy $E_{\rm HF}$ taken with the opposite sign differs from
the more accurate experimental value of the ionization potential
$I_{\rm exp}$. This correction is given by
\begin{equation}
A= A_{\rm HF} \, \left(\gamma_{\rm exp}/\gamma \right )^{Z_a/\gamma+1/2} \,,
\end{equation}
where $\gamma_{\rm exp}= \sqrt{2 \, I_{\rm exp}}$ \,.

The numerical values for $C_{n^*l}$ obtained within different approximations are presented in Table~\ref{tab:cnl}. We also calculate the relative error defined for a given $C_{n^*l}$ as $\Delta_{\text{rel}}=|C_{n^*l}^{\text{exact}}-C_{n^*l}|/C_{n^*l}^{\text{exact}}$. 
Note that the first value for $C_{n^*l}$ in the Hartree approximation (denoted with asterisk) is set manually, as Eq.~\eqref{hartree} is not applicable to neutral atoms [Eq.~\eqref{hartree} can take zero or even a complex value in this case]. 
This practical workaround allows using the Hartree approximation in PIC simulations. 
It should not affect the overall precision of calculations as long as strong fields and deep ionization states are considered.

	\begin{table}
            \caption{Coefficients $C_{n^*l}$ and relative error $\Delta_{\text{rel}}$ (with respect to the calculated within different approximations for argon. $C_{n^*l}^{\text{exact}}$ is extracted from the numerical solution to the Schrodinger equation for asymptotic electron states.}
            \label{tab:cnl}
            \centering
		\begin{tabularx}{\linewidth}{XXXXXXX}
        \hline\hline
            Ion charge & Config. & $C_{n^*l}^{\text{exact}}$ & $C_{n^*l}^{\text{Hartree}}$ & $\Delta_{\text{rel}}$ Hartree & $C_{n^*l}^{\text{ADK}}$ & $\Delta_{\text{rel}}$ ADK \\\hline
        0 & $3p^6$ & 1.001 & 1.0* & 0.001 & 1.016 & 0.015 \\
        1 & $3p^5$ & 0.914 & 0.426 & 0.534 & 0.869 & 0.05 \\
        2 & $3p^4$ & 0.842 & 0.534 & 0.366 & 0.724 & 0.14 \\
        3 & $3p^3$ & 0.876 & 0.571 & 0.349 & 0.616 & 0.297 \\
        4 & $3p^2$ & 0.772 & 0.581 & 0.247 & 0.513 & 0.335 \\
        5 & $3p^1$ & 0.689 & 0.574 & 0.167 & 0.432 & 0.373 \\
        6 & $3s^2$ & 0.954 & 0.906 & 0.051 & 0.428 & 0.551 \\
        7 & $3s^1$ & 0.88 & 0.861 & 0.022 & 0.376 & 0.573 \\
        8 & $2p^6$ & 0.652 & 0.513 & 0.214 & 0.764 & 0.172 \\
        9 & $2p^5$ & 0.641 & 0.53 & 0.174 & 0.733 & 0.143 \\
        10 & $2p^4$ & 0.633 & 0.542 & 0.144 & 0.707 & 0.115 \\
        11 & $2p^3$ & 0.646 & 0.551 & 0.148 & 0.685 & 0.06 \\
        12 & $2p^2$ & 0.634 & 0.56 & 0.117 & 0.658 & 0.039 \\
        13 & $2p^1$ & 0.623 & 0.567 & 0.09 & 0.634 & 0.018 \\
        14 & $2s^2$ & 1.043 & 1.021 & 0.021 & 0.624 & 0.402 \\
        15 & $2s^1$ & 1.018 & 1.011 & 0.008 & 0.6 & 0.411 \\
        16 & $1s^2$ & 1.003 & 0.994 & 0.009 & 1.005 & 0.002 \\
        17 & $1s^1$ & 1.003 & 1.0 & 0.003 & 1.0 & 0.003 \\
            \hline\hline
		\end{tabularx}
	\end{table}

As can be seen from the table, the relative error for the Hartree approximation is higher than that of ADK for the outer $3p$-shell electrons. However, for deeper shells the $\Delta_{\text{rel}}$ becomes smaller for the former. Recall that the overall precision of the tunneling formula \eqref{PPT_definition} is not that high for outer-shell electrons (see Section~\ref{sec:ii}). The lack of precision of Eq.~\eqref{hartree} for such states is less important than the limitations imposed by the sequential ionization approximation. Therefore, we conclude that as long as tunnel ionization of deeper shells (in practice, $Z>3$) is considered, the Hartree approximation is advantageous with respect to the ansatz used in the ADK. 
Hence, usage of Eq.~\eqref{hartree} in the PPT rate \eqref{PPT_definition} is preferred unless a better approximation is available.

As a practical test, we calculate the ion profiles as in Section~\ref{sec:vc} (assuming the sequential ionization order) with different values for $C_{n^*l}$. 
Results are presented in Fig.~\ref{fig:profiles_cnl}. 
In addition to the three sets of $C_{n^*l}$ listed in Tab.~\ref{tab:cnl}, we also test a ``simple man's'' approximation setting $C_{n^*l}=1$. 
As can be seen from the plots, the results are very close. 
A slight difference is apparent for the $3p$ and $3s$ state profiles. 
The strongest deviation is observed in the forth sub-plot, where the $2p$ state thresholds are shifted to the left. 
Overall, for practical applications, all the discussed approximations for $C_{n^*l}$ appear to be generally robust. 
Moreover, as is shown in the main body of the paper, different models for the treatment of the $m$-dependence of the rates and the presence of nonsequential ionization pathways may have a much higher impact on the output of calculations.

\section{Estimation of errors}
\label{app:errors}
\begin{figure}
    \centering
    \includegraphics[width=\linewidth]{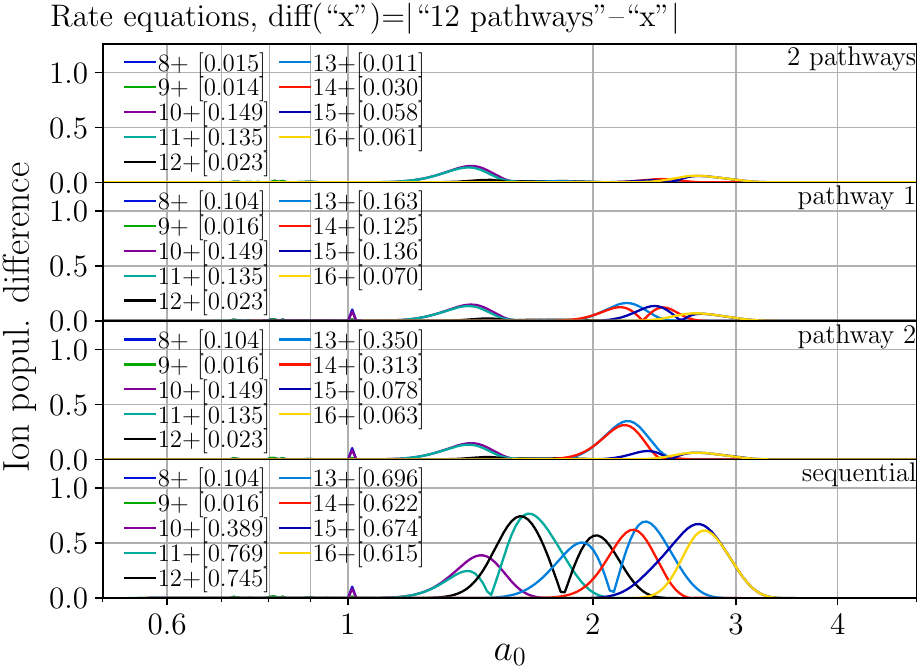}
    \caption{Absolute error for the ion populations presented in Fig.~\ref{fig:rate_eqs_profiles} (plots 2..5 are subtracted from the top plot therein by matching color). In the legend, numbers in brackets show the maximum absolute error.}
    \label{fig:errors_rate_eqs}
\end{figure}

\begin{figure}
    \centering
    $\,$\\[2ex]
    \includegraphics[width=\linewidth]{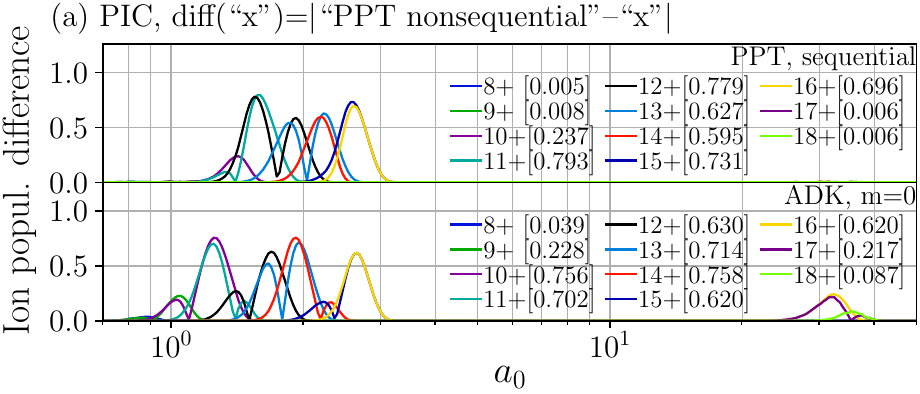}\\
    \includegraphics[width=\linewidth]{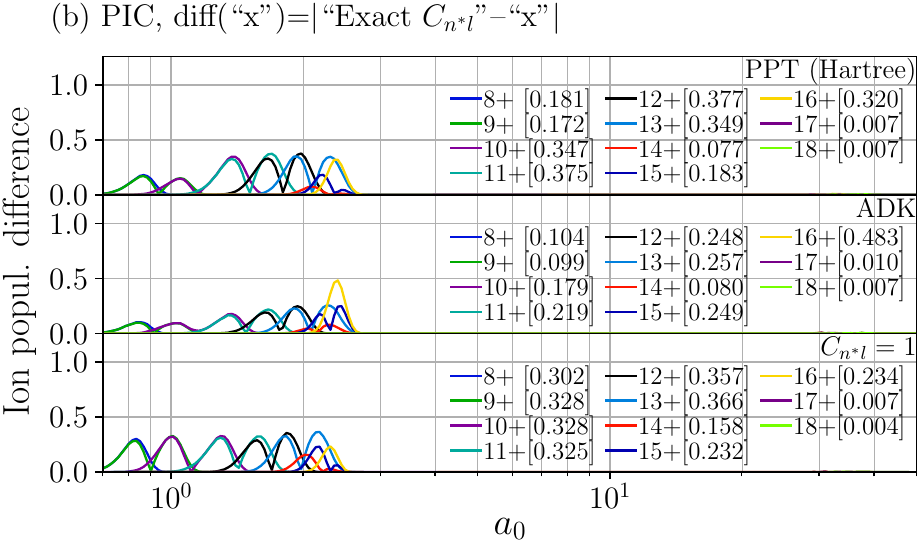}
    \caption{Absolute error for the ion populations presented in: (a) Fig.~\ref{fig:PIC_profiles} (plots 2 and 3 are subtracted from the top plot therein by matching color), (b) Fig.~\ref{fig:profiles_cnl} (plots 2..4 are subtracted from the top plot therein by matching color). In the legend, numbers in brackets show the maximum absolute error.}
    \label{fig:errors_PIC}
\end{figure}

In Figs.~\ref{fig:errors_rate_eqs} and \ref{fig:errors_PIC}, we provide additional data allowing for estimating the errors. Fig.~\ref{fig:errors_rate_eqs} shows the difference of the ion profiles calculated with the rate equations for different pathways, see Fig.~\ref{fig:rate_eqs_profiles}. 
In particular, for a given sub-plot in Fig.~\ref{fig:rate_eqs_profiles} (referred to by a label in the plot), we take a curve for each charge state and subtract it from the top plot ``12 pathways''. The correspondingly colored curve in Fig.~\ref{fig:errors_rate_eqs} shows the absolute difference. The numbers in the legend show the peak value of this difference for each charge state, namely, the maximum absolute error. One can see that the ``2 pathway'' approximation provides the smallest error overall, and ``pathway 1'' is the best one-pathway approximation for the full solution. Based on this, we pick ``pathway 1'' as the dominant pathway. We note that using ``pathway 2'' would still be a better precision approximation than sequential ionization. 

In the same spirit, we plot the ion profile difference for the models presented in Fig.~\ref{fig:PIC_profiles}. In Fig.~\ref{fig:errors_PIC}a, each plot corresponds to a subtraction of the respsectively labeled curves from the ``PPT, nonsequential'' model in Fig.~\ref{fig:PIC_profiles}. The same applies to Fig.~\ref{fig:errors_PIC}b, where we show the absolute error for the models presented in Fig.~\ref{fig:profiles_cnl} benchmarked against ``Exact $C_{n^*l}$.

\bibliography{lit_ionization_doi}

\end{document}